\documentclass[useAMS,usenatbib]{mn2e}
\usepackage{journal_names}
\usepackage{graphicx,times}
\usepackage{amsmath}
\usepackage[T1]{fontenc}
\usepackage{aecompl}
\usepackage{tabularx}
\usepackage{longtable}
\usepackage{pdflscape}
\usepackage{rotating}
\usepackage{multirow}

\def\la{\mathrel{\hbox{\rlap{\hbox{\lower4pt\hbox{$\sim$}}}\hbox{$<$}}}}
\def\ga{\mathrel{\hbox{\rlap{\hbox{\lower4pt\hbox{$\sim$}}}\hbox{$>$}}}}

\newcommand{\Lagr}{\mathcal{L}}
\def\apjl   {{ApJL }}
\def\apjs   {{ApJS }}

\title[Joint analysis of 6dFGS and SDSS peculiar velocities]{Joint analysis of 6dFGS and SDSS peculiar velocities for the growth rate of cosmic structure and tests of gravity}

\author[K.~Said et al.]
{\parbox{\textwidth}{Khaled~Said$^{1,2}$\thanks{E-mail: khaled.said@anu.edu.au},
Matthew Colless$^{1}$, Christina Magoulas$^{3}$, John R. Lucey$^{4}$, and Michael~J.~Hudson$^{5,6,7}$} 
\vspace{0.4cm}\\
\parbox{\textwidth}{
$^{1}$Research School of Astronomy and Astrophysics, Australian National University, Canberra, ACT 2611, Australia\\
$^{2}$School of Mathematics and Physics, University of Queensland, Brisbane, QLD 4072, Australia\\
$^{3}$Australian Synchrotron, 800 Blackburn Road, Clayton, VIC 3168, Australia\\
$^{4}$Centre for Extragalactic Astronomy, Durham University, Durham DH1 3LE, United Kingdom\\ 
$^{5}$Department of Physics and Astronomy, University of Waterloo, 200 University Avenue West, Waterloo, ON N2L 3G1 Canada\\
$^{6}$Waterloo Centre for Astrophysics, University of Waterloo, 200 University Avenue West, Waterloo, ON N2L 3G1 Canada\\
$^{7}$Perimeter Institute for Theoretical Physics, 31 Caroline St N, Waterloo, ON N2L 2Y5, Canada}}

\begin{document}

\date{Accepted XXXX. Received XXXX; in original form XXXX}

\pagerange{\pageref{firstpage}--\pageref{lastpage}} \pubyear{2019}

\maketitle

\label{firstpage}

\begin{abstract}
Measurement of peculiar velocities by combining redshifts and distance indicators is a powerful way to measure the growth rate of cosmic structure and test theories of gravity at low redshift. Here we constrain the growth rate of structure by comparing observed Fundamental Plane peculiar velocities for 15894 galaxies from the 6dF Galaxy Survey (6dFGS) and Sloan Digital Sky Survey (SDSS) with predicted velocities and densities from the 2M$++$ redshift survey. We measure the velocity scale parameter $\beta \equiv {\Omega_m^\gamma}/b = 0.372^{+0.034}_{-0.050}$ and $0.314^{+0.031}_{-0.047}$ for 6dFGS and SDSS respectively, where $\Omega_m$ is the mass density parameter, $\gamma$ is the growth index, and $b$ is the bias parameter normalized to the characteristic luminosity of galaxies, $L^*$. Combining 6dFGS and SDSS we obtain $\beta= 0.341\pm0.024$, implying that the amplitude of the product of the growth rate and the mass fluctuation amplitude is $f\sigma_8 = 0.338\pm0.027$ at an effective redshift $z=0.035$. Adopting $\Omega_m = 0.315\pm0.007$ as favoured by Planck and using $\gamma=6/11$ for General Relativity and $\gamma=11/16$ for DGP gravity, we get $S_8(z=0) = \sigma_8 \sqrt{\Omega_m/0.3} =0.637 \pm 0.054$ and $0.741\pm0.062$ for GR and DGP respectively. This measurement agrees with other low-redshift probes of large scale structure but deviates by more than $3\sigma$ from the latest Planck CMB measurement. Our results favour values of the growth index $\gamma > 6/11$ or a Hubble constant $H_0 > 70$\,km\,s$^{-1}$\,Mpc$^{-1}$ or a fluctuation amplitude $\sigma_8 < 0.8$ or some combination of these. Imminent redshift surveys such as Taipan, DESI, WALLABY, and SKA1-MID will help to resolve this tension by measuring the growth rate of cosmic structure to 1\% in the redshift range $0 < z < 1$.  
\end{abstract}


\section{Introduction}

There have been many efforts in the last two decades to test Einstein's general theory of relativity (GR), motivated by the discovery of the accelerating expansion of the universe \citep{1998AJ....116.1009R,1999ApJ...517..565P}. This cosmic acceleration can be explained within GR by invoking an appropriate value of Einstein's cosmological constant \citep{2001LRR.....4....1C,2003RvMP...75..559P}.  An alternative explanation is that this cosmic acceleration arises as a result of new gravitational physics \citep{2000PhLB..485..208D,2002PhLB..540....1F,2002hep.th....9227A,2003astro.ph..1510D,2004PhRvD..70d3528C,2009PhRvD..80b4037C}. Measuring the growth rate of cosmic structure is one observational way to distinguish GR from alternative gravity theories, because the expansion history of the universe affects the growth rate of large-scale structures; for a recent review see \citealt{2015APh....63...23H}.

On sufficiently large scales the matter distribution in the universe is effectively homogeneous and the expansion of the universe is effectively uniform. To first order, this means that the recession velocity of a low-redshift galaxy is directly proportional to its distance (the Hubble-Lemaitre law):
\begin{equation}
cz = H_0r,
\end{equation}
where $z$ is the redshift and $cz$ the recession velocity of a galaxy, $r$ is its distance, and $H_0$ is the Hubble parameter giving the present-day ($z$=0) expansion rate. Currently there is tension at the 4--5$\sigma$ level between direct local measurements of the expansion rate ($H_0=73.24\pm1.74$~km~s$^{-1}$~Mpc$^{-1}$; \citealt{2016ApJ...826...56R}) and the more precise but cosmology-dependent measurements from the cosmic microwave background radiation ($H_0=67.4\pm0.5$~km~s$^{-1}$~Mpc$^{-1}$; \citealt{2018arXiv180706209P}). 

On smaller scales, however, regions of high and low density form due to gravitational amplification of tiny perturbations in the density field emerging from the Big Bang. As a result, most of the galaxies in our Universe deviate slightly from the Hubble-Lemaitre law because they have peculiar velocities (i.e.\ velocities peculiar to themselves that are not part of the general 'Hubble flow') caused by local inhomogeneities in the mass distribution. Thus in general the relationship between redshift, Hubble velocity and peculiar velocity is
\begin{equation}
cz = H_0r+[v(\textit{\textbf{r}})-v(0)],
\end{equation}  
where $v(\textit{\textbf{r}})$ and $v(0)$ are the peculiar velocities along the line of sight of the galaxy and the observer.

In linear perturbation theory, where density fluctuations are small relative to the mean density, the density contrast $\delta$ is
\begin{equation}
\delta(\textit{\textbf{r}}) \equiv \frac{\rho(\textit{\textbf{r}})-\rho_0}{\rho_0} \ll 1
\end{equation}  
where $\rho(\textit{\textbf{r}})$ is the mass density field and $\rho_0$ is the mean mass density. In this linear regime the peculiar velocities are directly proportional to the gravitational acceleration \citep{1980lssu.book.....P,1993ppc..book.....P,1995PhR...261..271S} and are given by
\begin{equation}
v(r) = \frac{H_0f}{4\pi}\int d^3\textit{\textbf{r}}' \delta(\textit{\textbf{r}}') \frac{\textit{\textbf{r}}'-\textit{\textbf{r}}}{|\textit{\textbf{r}}'-\textit{\textbf{r}}|^3},
\label{v-d}
\end{equation}
where $f$ is the growth rate of the perturbations and $H_0$ drops out when using distances in km~s$^{-1}$. Equation~\ref{v-d} shows that by measuring and comparing the mass density and peculiar velocity fields it is possible to constrain the growth rate of the perturbations. The growth rate $f$ can be parameterized as a function of the mass density parameter $\Omega_m \equiv \rho_m/\rho_0$ (where $\rho_0$ is the critical density) and the growth index $\gamma$ (which is determined by the theory of gravity):
\begin{equation}
f(z) = \Omega_m(z)^{\gamma}
\end{equation}   

For the standard $\Lambda$CDM cosmological model (with flat geometry, a cosmological constant and cold dark matter), $\gamma = 6/11$ \citep{1998ApJ...508..483W,2005PhRvD..72d3529L}, while for alternative theories of gravity $\gamma$ takes on other values---e.g.\ in the Dvali-Gabadadze-Porrati model (DGP; \citealt{2000PhLB..485..208D}), $\gamma=11/16$ \citep{2007APh....28..481L}.

In practice what we are constraining from Equation~\ref{v-d} is the velocity scale $\beta \equiv{f(z)/b}$, which is a combination of the growth rate and the linear biasing parameter $b$ that is the ratio of the density fluctuations in galaxy number and the density fluctuations in total mass. Since the bias parameter is usually unknown, it is common to use the product of the growth rate and the root mean square density fluctuation within spheres of 8~$h^{-1}$~Mpc, $f\sigma_8 = \beta \sigma_{8,g}$, where $\sigma_{8,g}= b \sigma_8$ is the root mean square fluctuation in galaxy number within spheres of 8~$h^{-1}$~Mpc and can be measured from redshift surveys alone.

Peculiar velocities of galaxies can be measured {\em statistically} via the  redshift-space distortions (RSD; \citealt{1987MNRAS.227....1K}) or {\em directly} for each galaxy via redshift-independent distance indicators. \cite{1995PhR...261..271S} used the integral over the power spectrum to compare the density fluctuations to the velocity fluctuations within a sphere of radius $R$. The velocity field includes more contribution from large scales than the mass field because there are two fewer factors of wavenumber $k$ in the velocity power spectrum integral. Thus direct peculiar velocity measurements complement statistical RSD measurements because direct measurements are sensitive to lower $k$ (large scales, up to hundreds of Mpc) while RSD have better statistical power at larger $k$ (lower scales, down to tens of Mpc).  \cite{2014MNRAS.445.4267K} quantified the improvement possible when these two methods are combined compared to using RSD only (see also \cite{2017MNRAS.464.2517H} for a detailed theoretical explanation). Moreover, comparing direct peculiar velocity measurements to the density field derived from redshift surveys is less sensitive to cosmic variance, whereas RSDs are sensitive to cosmic variance because they depend entirely on the density field. Such a comparison between the peculiar velocity and density fields is the focus of this paper.

There are two broad classes of redshift-independent distance indicators: (i)~indicators that do not require a primary calibration and are mostly accessible only at small distances, such as Cepheid variables \citep{1969PASP...81..707F}, and (ii)~indicators that do require a primary calibration but are accessible at much larger distances, such as Type~Ia supernovae \citep[SNe~Ia;][]{2012MNRAS.420..447T,2017JCAP...05..015H,2017ApJ...847..128H,2019astro2020T.270S}, the Tully-Fisher relation for spiral galaxies \citep{1977A&A....54..661T}, and the Fundamental Plane relation for early-type galaxies \citep{1987ApJ...313...59D,1987ApJ...313L..37D}. Cepheid variables currently provide the most accurate measurements of distances on extragalactic scales, although they become very faint beyond $\sim$20~Mpc and so are not useful for large peculiar velocity surveys. SNe~Ia are the most precise of the second class of indicators and are accessible to very large distances, although they are rare and measuring their distances requires observations at multiple epochs. This situation will change soon, however, as LSST is expected to observe $\sim$10$^7$ SNe~Ia in the first ten years of its survey \citep{2009arXiv0912.0201L}. However, most of these will be at very large distance so will have large errors in km s$^{-1}$. Hence, for the time being, Tully-Fisher and Fundamental Plane distances are the workhorse methods to directly measure peculiar velocities for thousands of galaxies in the local universe ($z \leq 0.1$).

Here we briefly summarize recent peculiar velocity studies that are relevant to compare with our findings. We refer the reader to Table~3 of \cite{2005ApJ...635...11P} for a summary of results prior to 2005.  

\cite{2011MNRAS.413.2906D} used a set of 2830 spiral galaxies with $200<cz<10000$~km~s$^{-1}$ from the SFI$++$ survey \citep{2006ApJ...653..861M,2007ApJS..172..599S} and derived peculiar velocities for these galaxies using the inverse Tully-Fisher relation. They compared these peculiar velocities to the predicted velocities derived from the density field based on the 2MASS Redshift Survey (2MRS; \citealt{2005ASPC..329..135H}) and obtained $\beta =0.33\pm0.04$, combining their result with a value of $\sigma_{8,g}=0.97\pm0.05$ for 2MASS redshift survey calculated by \cite{2007PhDT.........3W} suggesting $f\sigma_8 = 0.31\pm0.06$. Assuming $\Omega_m = 0.266$ (\textit{WMAP}: \citealt{2011ApJS..192...16L}) and $\sigma_{8g} = 0.97\pm0.05$ \citep{2007PhDT.........3W}, they reported $S_8 = \sigma_8 \sqrt{\Omega_m/0.3} = 0.61\pm0.10$. Although this measurement is within $1.5\sigma$ of the \textit{WMAP} results, it favours a low value of $\sigma_8$, which agrees more with other low-redshift estimates based on large-scale structure. 

\cite{2012ApJ...751L..30H} calculated $f\sigma_8$ from various samples. Comparing peculiar velocities from \cite{2009MNRAS.392..743W} to the density field from the IRAS Point Source Catalog (PSCz; \citealt{2000MNRAS.317...55S}), they found $f\sigma_8 = 0.37\pm0.04$ after marginalizing over the external bulk flow. Combining \cite{2011MNRAS.413.2906D} and \cite{2012MNRAS.420..447T}, they found $f\sigma_8 = 0.36\pm0.04$ by averaging their results. They also used these peculiar velocity measurements to obtain $\Omega_m = 0.259\pm0.045$ and $S_8 = \sigma_8 \sqrt{\Omega_m/0.3} = 0.695\pm0.032$. In addition, they measured a growth index $\gamma = 0.619\pm0.054$. These results also favoured a lower density, lower $\sigma_8$, and higher growth index $\gamma$ than the standard values obtained from the Planck CMB measurements, although in agreement with all other low-redshift probes. 

\cite{2012MNRAS.425.2880M} compared measured peculiar velocities from different surveys such as ENEAR \citep{2000AJ....120...95D}, SN \citep{2003ApJ...594....1T}, SFI++ \citep{2007ApJS..172..599S}, and A1SN \citep{2012ApJ...751L..30H} to the velocity field predicted from PSCz \citep{2000MNRAS.317...55S}. They limited their comparison to objects within 70\,$h^{-1}$\,Mpc because at larger distances the PSCz model starts to be too sparsely sampled and errors on measured velocities become large. They used a Bayesian hyper-parameter comparison for each catalogue as well as a joint comparison. Their result for the joint comparison is $f\sigma_8 = 0.42\pm0.03$.

\cite{2015MNRAS.450..317C} used the estimated distances for a sample of 2662 spiral galaxies from SFI$++$ survey \citep{2007ApJS..172..599S} and SNe. They compared these distances with the reconstructed distances from the 2M$++$ survey \citep{2011MNRAS.416.2840L}. They found $f\sigma_8 = 0.401\pm0.024$ which is in tension with \cite{2011MNRAS.413.2906D}, although they used almost the same sample of peculiar velocities from SFI$++$. In contrast, this result is in agreement with \cite{2012MNRAS.420..447T} and \cite{2005ApJ...635...11P}.

All the above-mentioned studies used either SNe~Ia or the Tully-Fisher relation for spiral galaxies because, at redshifts as low as $z$=0.02, there are more spiral than elliptical galaxies. However, at slightly higher redshifts around $z$=0.1, because of complex observational constraints it is easier to observe more ellipticals than spirals. Combining Tully-Fisher and Fundamental Plane surveys, and so using both types of galaxies, can yield larger peculiar velocity samples at both low and high redshifts.

In this paper, we take advantage of the already existing data from the Sloan Digital Sky Survey imaging and spectroscopy (SDSS; \citealt{2000AJ....120.1579Y,2002AJ....124.1810S}) in the northern hemisphere and the spectroscopic 6dF Galaxy Survey (6dFGS; \citealt{2009MNRAS.399..683J}) in the southern hemisphere to select a sample of early-type galaxies at $z < 0.1$ for our peculiar velocity survey. We compare the inferred peculiar velocities from the Fundamental Plane relation for these early-type galaxies with the peculiar velocities predicted from the 2M$++$ \citep{2011MNRAS.416.2840L,2015MNRAS.450..317C} density field to constrain the growth rate of cosmic structure.

Except where otherwise stated, we assume a  $\Lambda$CDM cosmology with $\Omega_m = 0.3$, $\Omega_\Lambda = 0.7$ and $H_0 = 100h$~km~s$^{-1}$~Mpc$^{-1}$. 

This paper is organized as follows. We present the sample selection and demonstrate the consistency of the SDSS and 6dFGS data in Section~2. In Section~3, we fit the Fundamental Plane using a 3D Gaussian model. The method for fitting the Fundamental Plane parameters and the velocity field simultaneously is presented in Section~4. The results are discussed in the context of previous work in Section~5. We summarize our results in Section~6.

\section{Data}

In this paper, we use the Fundamental Plane relation for early-type galaxies to measure distances and peculiar velocities. As for other distance indicators, the Fundamental Plane uses distance-independent observables to predict a distance-dependent parameter, which can then be compared with the corresponding observable to derive a distance estimate. The Fundamental Plane relation used here has the form
\begin{equation}
    \log R_e = a \log \sigma_0 + b \log I_e + c
\end{equation}
where $R_e$ is the (distance-dependent) effective radius (in kpc), $\sigma_0$ is the (distance-independent) central velocity dispersion (in km~s$^{-1}$), $I_e$ is the (distance-independent) mean surface brightness within the angular effective radius (in L$_{\odot}$~pc$^{-2}$), and $a$, $b$ and $c$ are the coefficients of the Fundamental Plane.  

Note that we do not measure the (distance-dependent) physical effective radius $R_e$ but rather the (distance-independent) angular effective radius $\theta_e$. Converting from angular to physical radius requires the angular diameter distance to the galaxy. For fitting the Fundamental Plane parameters and the velocity field, this conversion is done using the (unknown) true distance as a free parameter in the likelihood, which we then marginalise over (see Section~4).  

The central velocity dispersion can be measured using spectroscopic data, while the effective radius and mean surface brightness require an imaging survey. In this section we present the sample selection algorithm as well as the procedure that used to derive the required parameters for the Fundamental Plane analysis.

\subsection{6dFGS}

One of the main goals of the 6dF Galaxy Survey (6dFGS) was to measure peculiar velocities for a sample of early-type galaxies in the southern hemisphere. Fundamental Plane data for a sample of $\sim$9000 early-type galaxies was compiled by \cite{2012MNRAS.427..245M} and \cite{2014MNRAS.443.1231C}, and used by \cite{2014MNRAS.445.2677S} to derive peculiar velocities. We briefly recapitulate here how this sample was selected and how the three physical Fundamental Plane parameters were derived.

The sample selection criteria for the 6dFGS peculiar velocity (6dFGSv) sample were as follows:
(1)~2MASS $J$-band total apparent magnitude $m_J$\,$\leq$\,13.65; (2)~Q-value of 3, 4 or 5, indicating a reliable redshift; (3)~spectral  signal-to-noise ratio SNR\,$>$\,5~\AA$^{-1}$; (4)~redshift in the range 0.01\,$\leq$\,$z$\,$\leq$\,0.055; (5)~spectral template match parameter $R$\,$\geq$\,8; (6)~velocity dispersion lower limit $\sigma$\,$\geq$\,112~km~s$^{-1}$; and (7)~visual classification of the image and the spectrum as an early-type galaxy.

\cite{2014MNRAS.443.1231C} give the Fundamental Plane observables $\theta_e$, $\sigma_0$ and $I_e$ for the galaxies in the 6dFGS peculiar velocity sample. The effective apparent radius $\theta_e$ for each galaxy was determined as the empirical half-light radius corrected for the effects of the image point spread function using a S\'ersic model fit (see \citealt{2014MNRAS.443.1231C}, Section~3.1).

The central velocity dispersion $\sigma_0$ was derived from the measured 6dFGS velocity dispersion $\sigma$ using a two-step method (\citealt{2014MNRAS.443.1231C}, Section~5.1). First, an empirical relation was used to convert the angular effective radius measured from near-infrared 2MASS \citep{2000AJ....119.2498J} observations to an optical effective radius. Then the dispersion measured in the 6dF fibre aperture was converted to a central dispersion using the \cite{1995MNRAS.276.1341J} formula
\begin{equation}
    \frac{\sigma_0}{\sigma} = \left( \frac{\theta_e/8}{\theta_{\rm ap}}\right)^{-0.04}
\label{apcor}
\end{equation}
where $\theta_e/8$ is the standard aperture size (one-eighth of the optical effective radius) and $\theta_{\rm ap}=3.35$~arcsec is the 6dF fibre radius.

The mean surface brightness $I_e$ was derived by \cite{2014MNRAS.443.1231C} using the 2MASS $J$-band total apparent magnitude $m_J$ and the $J$-band angular effective radius $\theta_e$ after applying a surface brightness dimming correction, a spectral $k$-correction for redshift and evolution, and a Galactic extinction correction; these corrections are discussed in \cite{2014MNRAS.443.1231C}, Section~5.2.

\subsection{SDSS}

\begin{figure*}
\begin{center}
\includegraphics[width=\textwidth]{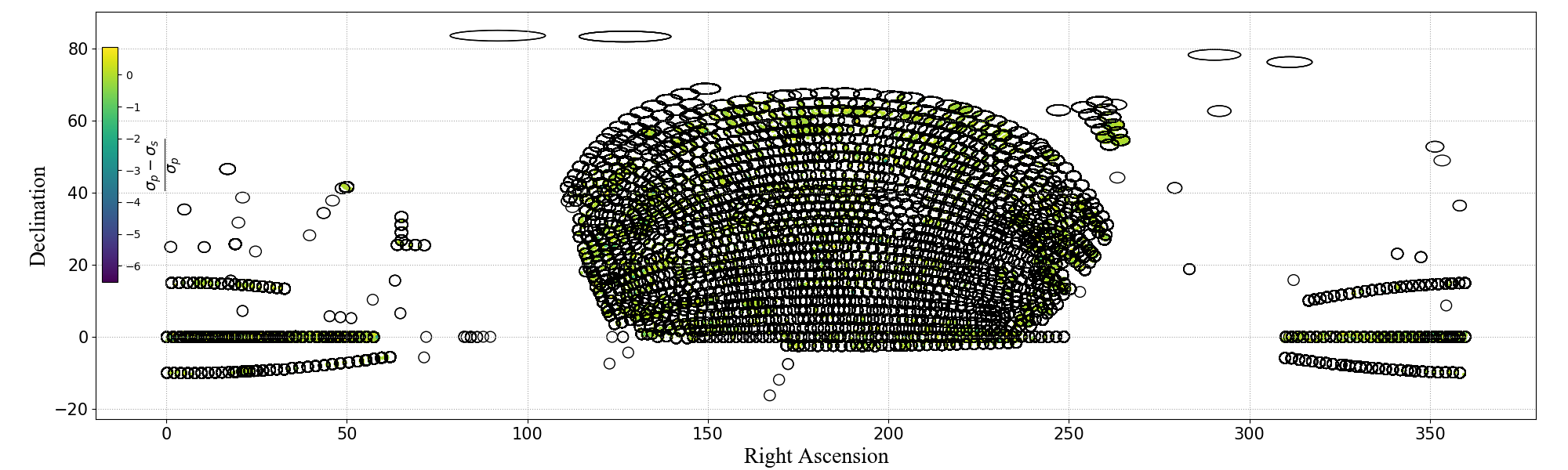}
\caption{The distribution of SDSS galaxies used in this paper to examine the consistency of the velocity dispersion measurements from plate to plate. The data covers $0^\circ$\,$\leq$\,RA\,$\leq$\,$360^\circ$, $-20^\circ$\,$\leq$\,Dec\,$\leq$\,$+90^\circ$, and 0.0033\,$\leq$\,$z$\,$\leq$\,0.3. Each solid circle represents a SDSS plate. Each dot shows a galaxy observed at least twice on different plates. Galaxies are colour-coded by the relative difference in velocity dispersion between the primary and secondary plates (scale given by the colour bar on the left-hand side).}
\label{overlap_regions_original_all}
\end{center}
\end{figure*}

The Sloan Digital Sky Survey (SDSS) data used in this paper were selected from SDSS Data Release~14 \citep[DR14;][]{2018ApJS..235...42A}. The selection criteria were chosen to provide a reliable sample of early-type galaxies with well-measured redshifts and velocity dispersions, and were as follows: (1)~$r$-band de~Vaucouleurs magnitude in the range 10.0\,$\leq$\,$m_r$\,$\leq$\,17.0; (2)~a reliable redshift measurement; (3)~spectrum classified as a galaxy; (4)~redshift in the range 0.0033\,$\leq$\,$z$\,$\leq$\,0.1; (5)~concentration index $r_{90}/r_{50}$\,$\geq$\,2.5 in $r$ and $i$ bands; (6)~likelihood of de~Vaucouleurs fit greater than likelihood of exponential fit in $r$ and $i$ bands; (7)~axial ratio $b/a$\,$\geq$\,0.3 in $r$ and $i$ bands; (8)~colour cut $g-r \geq 0.73 - 0.02(M_r + 20)$ \citep{2010MNRAS.405..783M}\footnote{We use $h$=0.7 to scale our cosmology to that of \citet{2010MNRAS.405..783M}.}; (9)~velocity dispersion lower limit of $\sigma$\,$\leq$\,70~km~s$^{-1}$; and (10)~no H$\alpha$ emission, EW$_{H\alpha}$\,$\geq$\,$-$1 (n.b.\ emission defined to be negative).

For galaxies in groups or clusters, we use the redshift of that group or cluster provided by \cite{2012A&A...540A.106T} instead of the redshift of the individual galaxy. This was also done for the 6dFGSv sample. For the purpose of constructing the Fundamental Plane parameters, we derived the angular effective radii $\theta_e$ from the angular de~Vaucouleurs fit scale radius $r_{\rm dev}$ using
\begin{equation}
    \theta_e = r_{\rm dev} \sqrt{b/a}
\end{equation}
where $b/a$ is the de~Vaucouleurs axial ratio. We subsequently converted the angular effective radius into physical effective radius by using the angular diameter distance \citep{1972gcpa.book.....W} corresponding to the observed redshift in the CMB frame and assuming our standard $\Lambda$CDM cosmology. Again, this is not what we will be using in Section~4. 

We calculated the effective surface brightness $\mu_e$ using the $r$-band de~Vaucouleurs apparent magnitude $m_r^{\rm  dev}$ as:
\begin{multline}
    \mu_e = m_r^{\rm  dev} + 0.85z + 2.5\log(2\pi \theta_e^2) \\ 
            - 2.5\log(1+z)^4 - k_r - A_r
\end{multline}
where $0.85z$ is the SDSS $r$-band evolution correction \citep{2003AJ....125.1849B}, $\theta_e$ is the angular effective radius in arcsec, \mbox{$2.5\log(1+z)^4$} is the surface brightness dimming correction, $k_r$ is the analytical approximation of the $K$-correction in the $r$-band given by \cite{2010MNRAS.405.1409C}, and $A_r$ is the Galactic extinction from \cite{2011ApJ...737..103S}. We converted the effective surface brightness $\mu_e$ from magnitude units to log-luminosity units using
\begin{equation}
    \log I_e = 0.4 M_{\odot}^{\lambda} - 0.4\mu_e +2 \log(206265/10)
\end{equation}
where $M_{\odot}^{\lambda}$ is the wavelength-dependent value for the absolute magnitude of the Sun \citep{2018ApJS..236...47W}. 

Most previous works in this field have reported systematic offsets between velocity dispersion measurements from different surveys (e.g.\ \citealt{1995ApJS..100..105M,1997MNRAS.291..461S,2000MNRAS.313..469S,1999MNRAS.305..259W,2001MNRAS.327..265H}). Although the major contribution of these systematic offsets comes from using different instruments, other non-negligible contributions are still not fully understood \citep{2000MNRAS.313..469S,1999MNRAS.305..259W}. Understanding and removing these systematic offsets is crucial because they can artificially generate false peculiar velocities. For a single-instrument survey like SDSS, the main sources of systematic field-to-field (or plate-to-plate) offsets, are likely to be variations in the observing conditions, changes in the instrumental setup, or changes in the data reduction.  

For the purpose of an internal consistency check of our SDSS Fundamental Plane sample, we selected galaxies using a more relaxed redshift cut (0.0033\,$\leq$\,$z$\,$\leq$\,0.3) than our Fundamental Plane sample. For each of these galaxies we checked whether: (i)~it has been observed two or more times using different plates (a `primary' plate with the highest overall SNR and one or more `secondary' plates with lower overall SNR); (ii)~the velocity dispersion measurement (from the primary plate) is above the instrumental resolution limit, $\sigma$\,$\leq$\,70~km~s$^{-1}$; and (iii)~the seeing during the exposure is measured, seeing$_{50}$\,$\neq$\,0 (from spPlate header). 

\begin{figure}
\begin{center}
\includegraphics[width=\columnwidth]{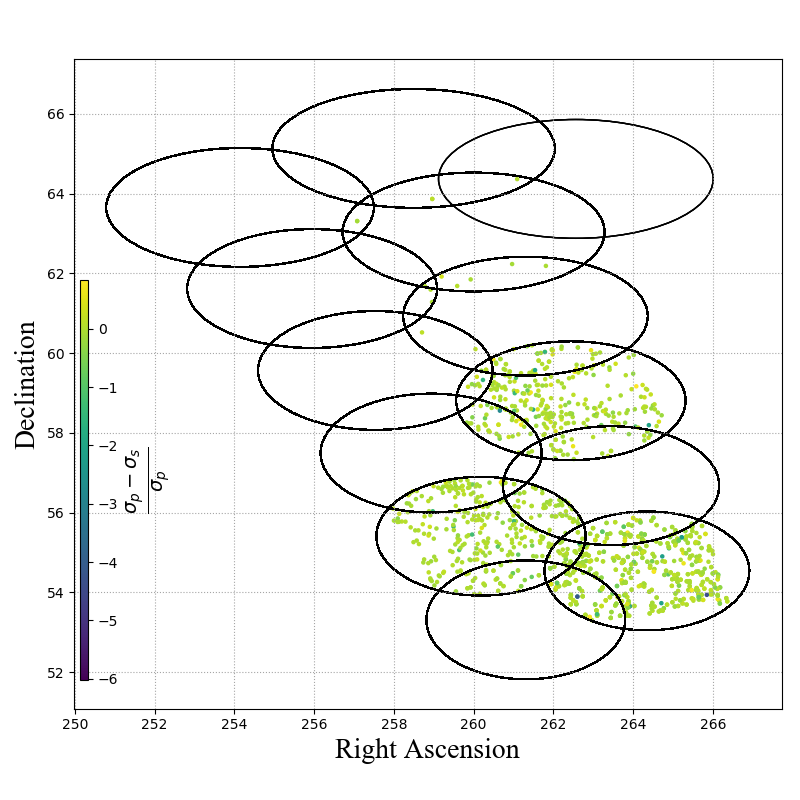}
\caption{Zoom-in on a tiled region in the SDSS survey. As in Figure~\ref{overlap_regions_original_all}, each circle is a SDSS plate and each dot represents a galaxy. Each of these galaxies has been observed more than once using different plates. In some cases the two overlapping plates are placed exactly on each other, although the observations were done on another date or different conditions. Galaxies are colour-coded by the relative difference in velocity dispersion measurements (see the colour scale at left).}
\label{overlap_regions_original_all_ra_dec}
\end{center}
\end{figure}

Figure~\ref{overlap_regions_original_all} shows the distribution of the sample used in this internal consistency check of the velocity dispersion measurements. Each plate used is shown as a black circle (primary as well as secondary) and each galaxy observed on more than one plate is shown as a dot colour-coded by the relative offset between the velocity dispersion measurements from the primary and secondary plates. Empty plates are the ones with galaxies observed more than once that did not pass the other selection criteria. The scale given by the colour bar is large and suggests that secondary plates occasionally have much larger velocity dispersions than the primary plates. However, Figure~\ref{overlap_regions_original_all_ra_dec}, a zoom-in example of a tiling region, shows that most galaxies have similar primary and secondary plate velocity dispersion measurements and only a few galaxies have large spurious dispersion differences; these are excluded from the analysis (see below).  

Figure~\ref{overlap_regions_original_all_ra_dec} also shows that some overlapping plates have the exact same position, resulting in a large number of galaxies in common; other plates overlap only partially, but these have the power to tie together different plates and so calibrate the whole survey to a common velocity dispersion scale. 

We used the relative error between pairs of observations to quantitatively check the consistency of the velocity dispersion measurements and test whether there are systematic offsets between observations/plates in the SDSS data. We define the pairwise relative error as
\begin{equation}
    \epsilon = \frac{\sigma_p-\sigma_s}{(\Delta \sigma_p^2 + \Delta \sigma_s^2)^\frac{1}{2}}
\end{equation}
where $\sigma_p$, $\sigma_s$, $\Delta \sigma_p$, and $\Delta \sigma_s$ are the velocity dispersion measurements from primary and secondary plates along with their associated errors, respectively. Consistent and unbiased velocity dispersion measurements with correctly estimated errors should give a Gaussian with a mean of zero and a standard deviation of unity. 

The sample of galaxies with overlap velocity dispersion measurements contains 2403 measurement pairs for 2102 individual galaxies. We remove extreme outliers by applying $3.5\sigma$ clipping, which excluded 34 measurements (1.4\%), leaving a final sample of 2369 measurements for 2069 galaxies. 

\begin{figure}
\begin{center}
\includegraphics[width=\columnwidth]{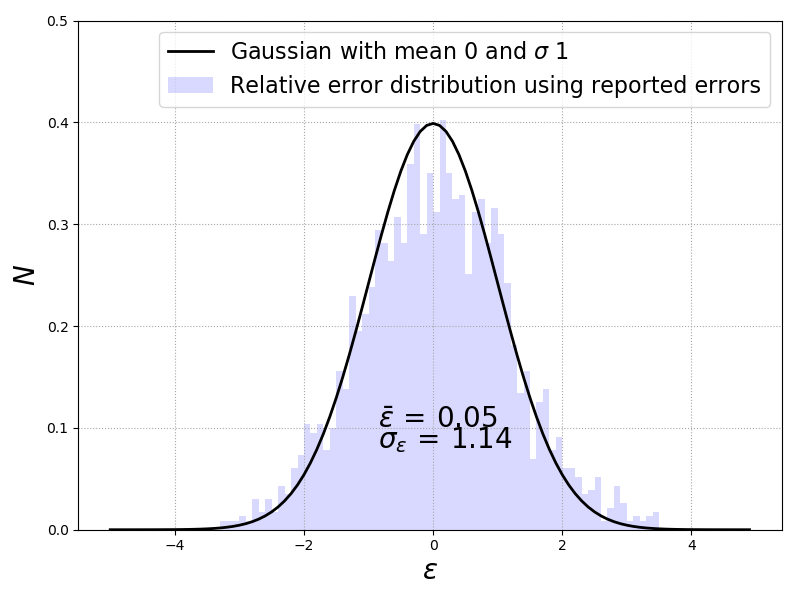}
\includegraphics[width=\columnwidth]{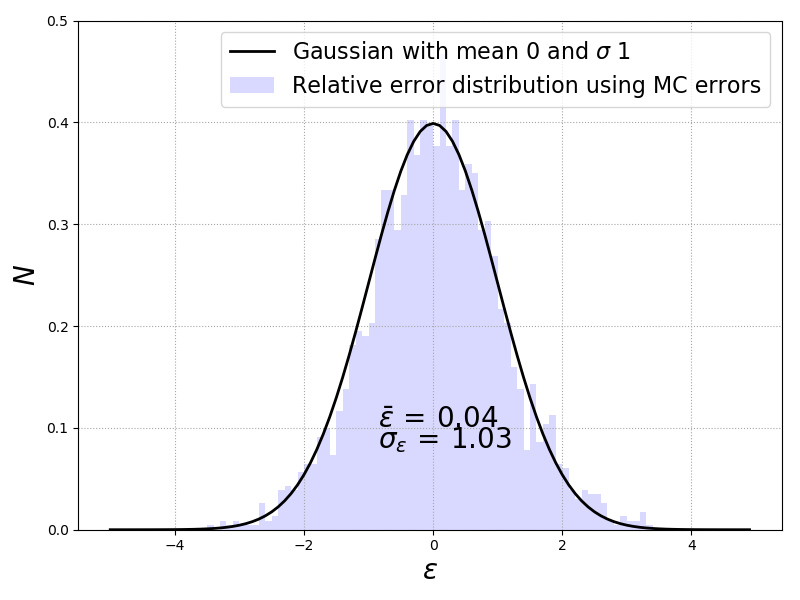}
\caption{Histograms of pairwise relative errors, $\epsilon$, in velocity dispersion measurements. The top panel shows measurements from SDSS DR14; the bottom panel shows measurements using pPXF and Monte Carlo error estimates. Consistent measurements and errors would produce a Gaussian with mean of zero and standard deviation of unity, as shown by the solid curves.}
\label{pull-distribution}
\end{center}
\end{figure}

The top panel of Figure~\ref{pull-distribution} shows the distribution of pairwise relative errors for the velocity dispersion measurements in SDSS DR14. The mean of the distribution is $\Bar{\epsilon} = 0.05$ and the width $\sigma_\epsilon = 1.14$; we over-plot a Gaussian with a mean of zero and standard deviation of unity for comparison. The offset in the mean is less than 2.5~times the standard error in the mean, and so not significant; however the offset from unity of the standard deviation is 9~times the uncertainty in the standard deviation, $\sigma/\sqrt{(2N-2)} = 0.016$, and so highly significant. There is also a marked flattening of the observed distribution around the peak.

We used the Pearson sample correlation coefficient $r$ to check if these differences between the primary and the secondary plates correlate with observational conditions. We looked for correlations of observing parameters with the velocity dispersion $\sigma_p$, the difference in the velocity dispersion measurements $\sigma_p-\sigma_s$, the fractional difference $(\sigma_p-\sigma_s)/\sigma_p$ and the velocity dispersion error $\Delta\sigma$. We found that higher SNR correlates with smaller velocity dispersion error, as expected, and with higher dispersion (presumably because higher dispersion correlates with higher luminosity and SNR). However we found no significant correlations with seeing or position on the sky.



The data reduction pipeline and the choice of spectral templates can also contribute to the difference in the velocity dispersion measurements $\sigma_p-\sigma_s$. To explore this potential issue, we re-measured the velocity dispersion of all galaxies in the sample, both from primary and secondary plates, using the penalised PiXel Fitting code (pPXF; \citealt{2004PASP..116..138C,2017MNRAS.466..798C}) and the full MILES stellar template library \citep{2006MNRAS.371..703S,2011A&A...532A..95F}. The associated error was calculated using Monte Carlo estimation, which gives typical errors of 6--7\% compared to the 4--5\% reported by the SDSS pipeline at this low redshift ($z$$<$0.1). 

The bottom panel of Figure~\ref{pull-distribution} shows the distribution of relative errors for these new measurements. Now the mean of the distribution is $\Bar{x} = 0.04$ and the width $\sigma=1.03$. Although the new measurements do not change the mean of the distribution, which is consistent with zero in both cases, they narrow the width by 0.11, bringing the standard deviation close to unity. This reduction of the width is mainly due to the larger error estimates for these measurements compared to the SDSS pipeline reported errors. The new measurements also remove the inconsistency near the peak of the Gaussian that is apparent in the SDSS pipeline measurements. 

\begin{figure}
\begin{center}
\includegraphics[width=\columnwidth]{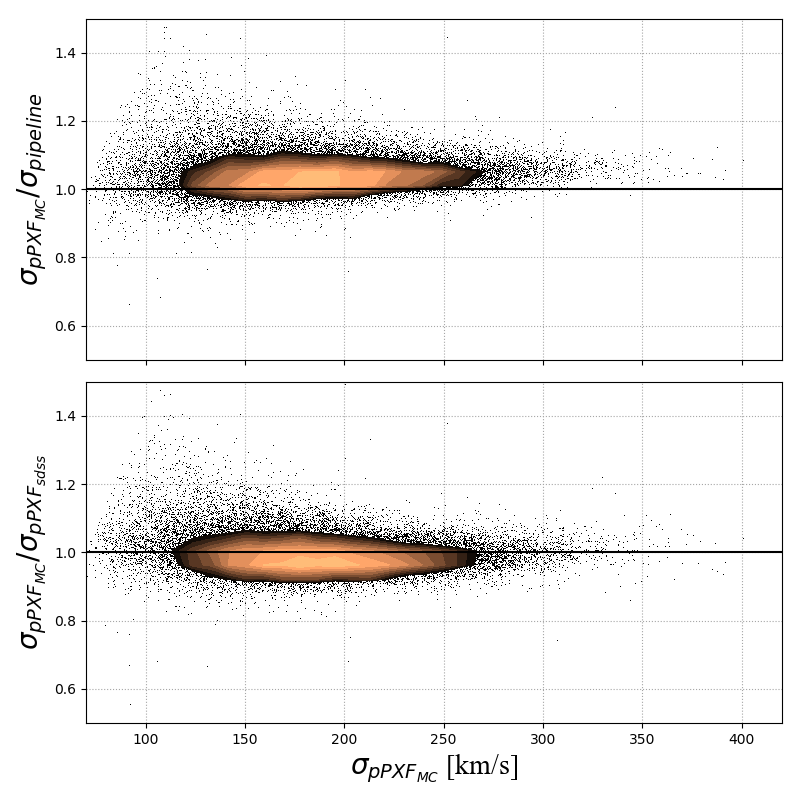}
\caption{Comparison of velocity dispersion measurements. The top panel compares the velocity dispersions measured by us using pPXF and Monte Carlo error estimates ($\sigma_{\textit{pPXF}_{\rm MC}}$) to the velocity dispersions reported by the SDSS pipeline ($\sigma_\textit{pipeline}$). There is an offset over the whole range, with our measurements slightly higher than those reported by the SDSS pipeline. The bottom panel compares the velocity dispersions measured by us using pPXF and Monte Carlo error estimates ($\sigma_{\textit{pPXF}_{\rm MC}}$) to the velocity dispersions reported by SDSS using pPXF ($\sigma_{\textit{pPXF}_{\rm SDSS}}$). In this case there is better agreement between the two measurements. Contours indicate the density of galaxies in both panels.}
\label{sigma_ppxf_sdss_vs_ratio_mc_to_ppxf_01_both}
\end{center}
\end{figure}

These improvements motivated us to re-measure the velocity dispersions using this method for our SDSS Fundamental Plane sample. Figure~\ref{sigma_ppxf_sdss_vs_ratio_mc_to_ppxf_01_both} shows two comparisons: first, we compare our new velocity dispersion measurements using pPXF and the full MILES library to the reported SDSS DR14 velocity dispersions; second, we compare our measurements to another set of velocity dispersion measurements reported in the galaxy properties catalog from the Portsmouth group (emissionLinesPort). In this figure our velocity dispersion measurement is $\sigma_{\textit{pPXF}_{\rm  MC}}$, the SDSS pipeline velocity dispersion is $\sigma_{\textit{pipeline}}$, and the Portsmouth group velocity dispersion is $\sigma_{\textit{pPXF}_{\rm  SDSS}}$. The Portsmouth group used the Gas AND Absorption Line Fitting (GANDALF; \citealt{2006MNRAS.366.1151S}) and pPXF packages with a set of model spectra from \cite{2011MNRAS.418.2785M, 2011MNRAS.412.2183T} based on the MILES library. There is an offset between our measurements and the ones reported by the SDSS pipeline, with our measurements being slightly higher. However, there is better agreement with the Portsmouth group's velocity dispersions, part of which is due to using similar methods.  

We corrected the fibre velocity dispersions we measured for SDSS galaxies to central velocity dispersions using Equation~\ref{apcor}, where $\theta_{ap}=1.5$~arcsec is the SDSS fibre radius. Throughout the rest of this paper we use only these new velocity dispersion measurements for the SDSS galaxies.

\section{Fundamental Plane fits}
\label{FPfits}
The main goal of this section is to test and compare our new SDSS data set on a well established and previously used method. For comparison with previous work, we used a 3D Gaussian model to fit the Fundamental Plane. This method was first proposed by \cite{2001MNRAS.321..277C} to measure the peculiar velocities of early-type galaxies and we refer the reader to \cite{2012MNRAS.427..245M} for a full explanation of the details. In brief, we first define all Fundamental Plane quantities in logarithmic units: $r = \log R_e$, $s = \log \sigma_0$, and $i=\log I_e$. Assuming that the joint distribution of these three quantities is well represented by a Gaussian, the three-dimensional probability distribution in $(r,s,i)$ space for any galaxy, $n$, is defined as
\begin{equation}
    P(x_n) = \frac{\exp[{-0.5\textbf{x}_n^T (\textbf{V} + \textbf{E}_n)^{-1} \textbf{x}_n}]}
                  {(2\pi)^{3/2}|\textbf{V}+\textbf{E}_n|^{1/2}f_n}
\end{equation}
where $\textbf{x}_n = (r-\Bar{r},s-\Bar{s},i-\Bar{\imath})$ is the position of galaxy $n$ in the Fundamental Plane space, $\textbf{V}$ is the variance matrix that defines the intrinsic scatter of the Fundamental Plane as, 
\begin{equation}
    \textbf{V} = \begin{pmatrix}
    \sigma_1^2 & 0 & 0\\
    0 & \sigma_2^2 & 0\\
    0 & 0 & \sigma_3^3
    \end{pmatrix},
\end{equation}

$\textbf{E}_n$ is the error matrix of the observables for galaxy $n$ defined as,

\begin{equation}
    \textbf{E} = \begin{pmatrix}
    \epsilon_{r}^{2} + \epsilon_{rp}^{2} & 0 & \rho_{ri} \epsilon_{r} \epsilon_{i}\\
    0 & \epsilon_{s}^{2} & 0 \\
    \rho_{ri} \epsilon_{r} \epsilon_{i} & 0 & \epsilon_{i}^{2}
    \end{pmatrix},
\end{equation}
where $\epsilon_r$, $\epsilon_s$, and $\epsilon_i$ are the errors on the Fundamental Plane parameters $r$, $s$, and $i$, respectively. The conversion from angular to physical radius assumes that each galaxy has zero peculiar velocity (i.e.\ using the redshift as distance). We account for this through an additional error in the observational error matrix $\epsilon_{rp} = \log(1+300/cz)$, which assumes a peculiar velocity of 300\,km\,s$^{-1}$ for every galaxy in the sample \citep{1995PhR...261..271S}. The correlation between errors in $r$ and $i$ is accounted for using the correlation coefficient $\rho_{ri}$, which was found to be 0.95 for the 6dF sample and 1.0 for the SDSS sample. $f_n$ is the normalization factor that accounts for the selection cuts and makes the integral over the probability distribution equal to unity, $\int P(x) d^3x = 1$. 

We then maximize the sample likelihood to determine the Fundamental Plane parameters: the mean values $\Bar{r}$, $\Bar{s}$, $\Bar{\imath}$, and the variance matrix $\textbf{V}$. The likelihood is
\begin{equation}
    \Lagr = \prod_{n=1}^{N_g} P(x_n)^{1/S_n},
\end{equation}
where $1/S_n$ is a weighting factor applied to each galaxy according to the sample selection function (the fraction of galaxies with observed parameters similar to galaxy $n$ that are included in the sample). Each galaxy in the sample is thus treated in the fitting procedure as $1/S_n$ galaxies. 

For both 6dFGS and SDSS samples, the selection function depends on apparent magnitude. For each galaxy $n$, the selection probability is defined as:
\begin{equation}
    S_n=\left\{
                \begin{array}{l r}
                  1 \; \; \; &  \; z_n^{\rm  max} \geq z_{\rm max}\\
                  \displaystyle\frac{V_n^{\rm max}-V(z_{\rm min})}{V(z_{\rm max})-V(z_{\rm min})} &  z_{\rm min} < z_n^{\rm max} < z_{\rm max}\\
                  0 \; \; \; & \; z_n^{\rm max} \leq z_{\rm min}
                \end{array}
              \right.
\end{equation}
where $z_{\rm min}$ and $z_{\rm max}$ are the upper and lower redshift limits for the survey and $V(z_{\rm min})$ and $V(z_{\rm max})$ the corresponding comoving volumes. Similarly, $z_n^{\rm  max}$, and $V_n^{\rm max}$ are the maximum redshift and comoving volume to which galaxy $n$ can be detected given the survey apparent magnitude limit.

We did not carry out this analysis again on the 6dFGS sample as this has already been done twice, by \cite{2012MNRAS.427..245M} and \cite{2014MNRAS.445.2677S}. However, we apply it, for the first time, to the SDSS $r$-band sample, which contains $24848$ galaxies with redshift $z<0.1$. The best-fit Fundamental Plane parameters derived from this analysis of the SDSS $r$-band sample, along with those from the 6dFGS sample obtained by \cite{2012MNRAS.427..245M}, are given in Table~\ref{3D_gaussian_6df_sdss}. 

\begin{table}
\begin{center}
\caption[]{3D Gaussian fit for 6dF $J$-band, and SDSS $r$-band Fundamental Plane.}
\begin{tabular} { c  c   c}
\hline\hline
Parameter &  6dFGS \citep{2012MNRAS.427..245M} &  SDSS (this work) \\
\hline
$N_g$         & 8803               & 24848 \\
$a$            &$~~1.523\pm0.026$ &$~~1.461\pm0.014$\\
$b$            & $-0.885\pm0.008$ & $-0.822\pm0.0005$\\
$c$            & $-0.330\pm0.054$ & $-0.841\pm0.026$\\
$\Bar{r}$      &$~~0.184\pm0.004$ &$~~0.158\pm0.004$\\
$\Bar{s}$      &$~~2.188\pm0.004$ &$~~2.213\pm0.002$\\
$\Bar{\imath}$ &$~~3.188\pm0.004$ &$~~2.717\pm0.004$\\
$\sigma_1$     &$~~0.053\pm0.001$ &$~~0.0509\pm0.0004$\\
$\sigma_2$     &$~~0.318\pm0.004$ &$~~0.403\pm0.004$\\
$\sigma_3$     &$~~0.170\pm0.003$ &$~~0.195\pm0.002$\\
\hline
\end{tabular}
\label{3D_gaussian_6df_sdss}
\end{center}
\end{table} 

\begin{figure*}
\begin{center}
\includegraphics[width=\textwidth]{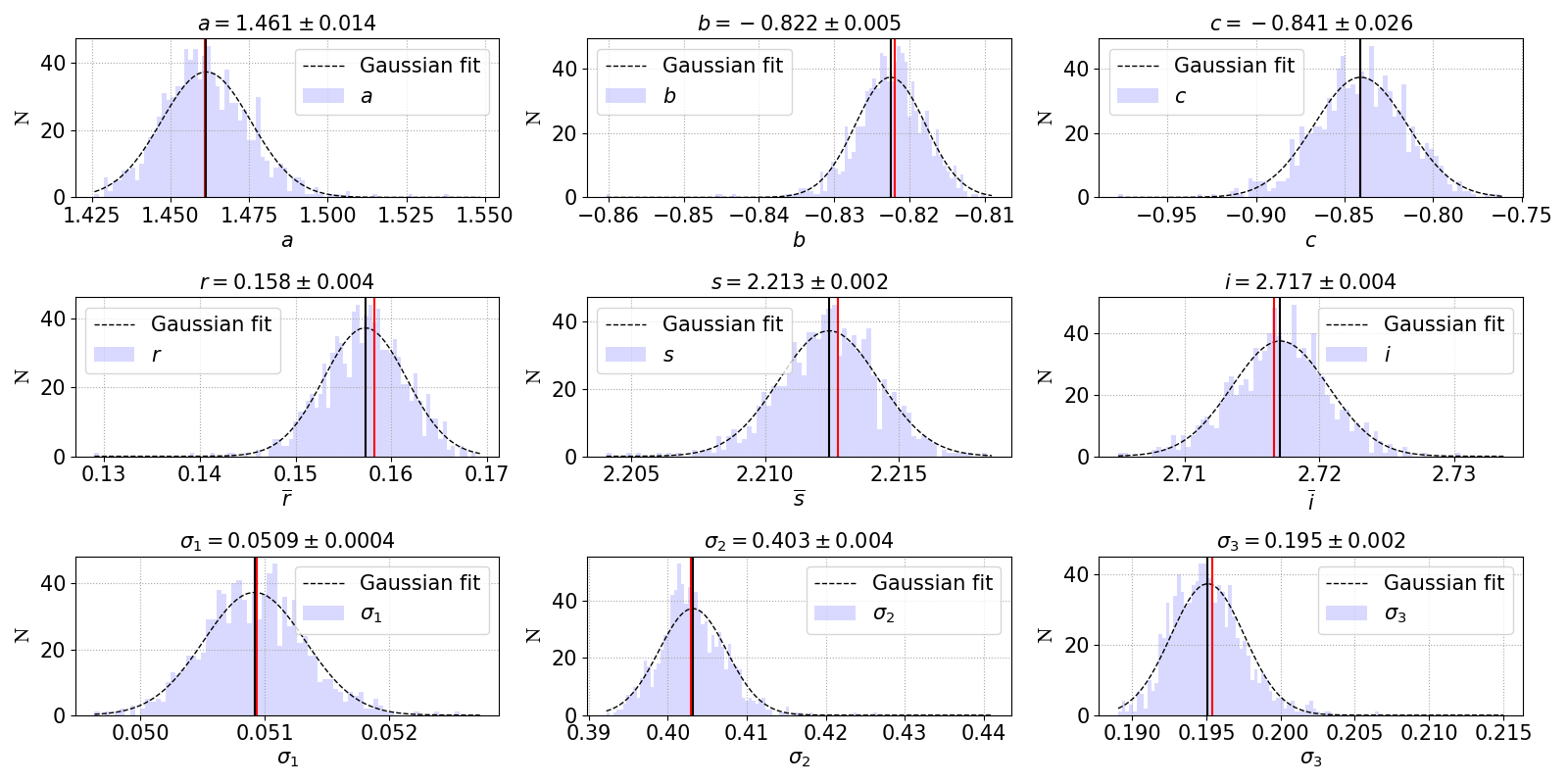}
\caption{The distribution of SDSS $r$-band Fundamental Plane parameters $a$, $b$, $c$, $\Bar{r}$, $\Bar{s}$, $\Bar{\imath}$, $\sigma_1$, $\sigma_2$, $\sigma_3$ derived from 1000 simulations of the best-fitting 3D Gaussian Fundamental Plane. The red vertical line in each panel shows the best-fit parameter value used to generate the mock, while the black vertical line shows the mean fitted value from the 1000 mocks. The best-fit Gaussian to the distribution of derived parameters is shown by the dashed curve. The input parameters to generate the mocks are shown at the top of each panel along with the RMS errors from the mocks.}
\label{Fundamental Plane_parameters}
\end{center}
\end{figure*}

Figure~\ref{Fundamental Plane_parameters} shows the error distributions of the Fundamental Plane parameters derived by fitting a 3D Gaussian model to each of 1000 mock SDSS samples. Each mock sample has 24848 mock galaxies (the same number as the SDSS $r$-band sample), which are drawn from a 3D Gaussian model for the Fundamental Plane with the same parameters as the best fit to the SDSS $r$-band sample. We used the same mock sample algorithm proposed and used by \cite{2012MNRAS.427..245M}. These mock samples were designed to be robust and well calibrated, as they serve different purposes (cf.\ \citealt{2012MNRAS.427..245M}, section~4). They have been extensively used to compare different fitting algorithms, to perform full validation tests of the fitting methods, to correct for any biases, and to define the accuracy and the precision of the fits. We refer the reader to Section~4 in \cite{2012MNRAS.427..245M} for a detailed description of the algorithm for generating mock samples and the functions they serve. 

Here we give the key steps in the mock sample algorithm: (1)~randomly generate \textbf{\textit{v}}-space variables with the corresponding variance \textbf{V}, then transform them to give $r$, $s$ and $i$ for a mock galaxy using the given values of $a$, $b$, $\Bar{r}$, $\Bar{s}$ and $\Bar{i}$; (2)~randomly generate a comoving distance from a uniform density distribution within $z<0.1$ and the assumed cosmology, then use this to convert from angular to physical radius and to give the redshift of the mock galaxy; (3)~calculate the apparent magnitude using the surface brightness and effective radius, then use the estimated uncertainties based on this magnitude to randomly generate Gaussian measurement errors in $r$, $s$ and $i$; (4)~use the derived errors to obtain the observed values of $r$, $s$ and $i$, and compute the observed magnitude for the mock galaxy using these values; finally, (5)~compute the selection probability for the mock galaxy.

\begin{figure*}
\begin{center}
\includegraphics[width=\textwidth]{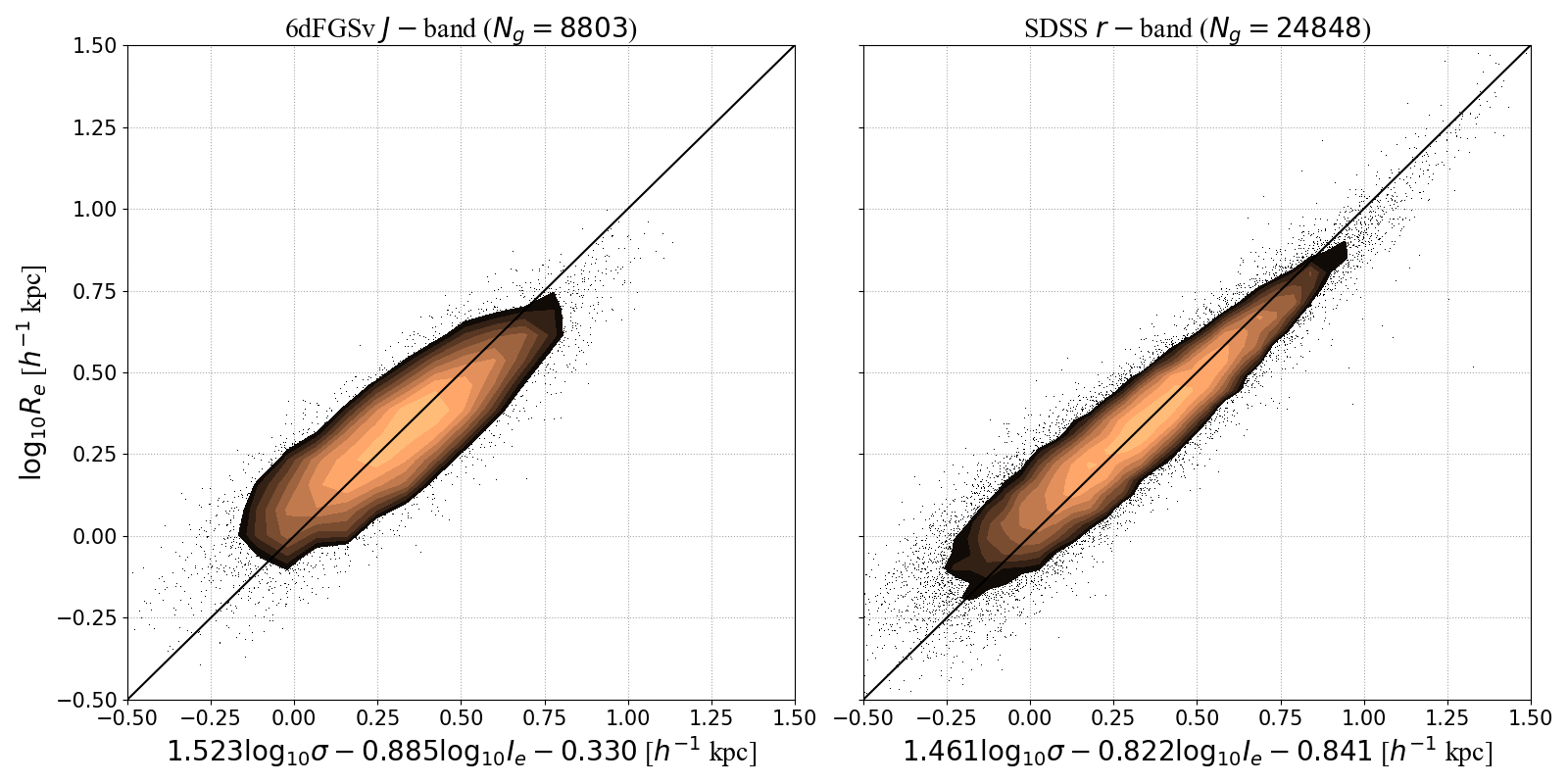}
\caption{The projected Fundamental Planes of 6dFGSv $J$-band (left-panel) and SDSS $r$-band (right-panel). The solid black line is the one-to-one line.}
\label{Forward_6df_sdss}
\end{center}
\end{figure*}

In each panel of Figure 5, the solid red line shows the fitted value of the parameter derived from the SDSS sample and the histogram is the distribution of the fitted parameter from the 1000 mocks. The dashed curve is the Gaussian fit to this distribution; the solid black line is the mean and the RMS provides the estimated error on the parameter. Figure~\ref{Forward_6df_sdss} shows the projected forward FP for both 6dFGSv $J$-band (left-panel) and SDSS $r$-band (right-panel).   

Because there are wavelength dependent FP tilts and offsets, one does not expect the FP coefficients to be identical because 6dF is $J$-band and SDSS is $r$-band. The best way to compare the Fundamental Plane fits to the 6dFGS and SDSS samples given in Table~\ref{3D_gaussian_6df_sdss} is by using the RMS scatter of the Fundamental Plane in the $r$  direction, which is directly proportional to the true distance error. The true distance error depends on additional factors such as the bias correction and the distribution of galaxies in the Fundamental Plane \citep{2012MNRAS.427..245M,2014MNRAS.445.2677S}.

We calculated the total RMS scatter in $r$ as:
\begin{equation}
    \sigma_r = [(a\epsilon_s)^2 + \epsilon^2_{\rm  phot} + \sigma^2_{r,{\rm int}}]^{1/2}
\end{equation}
where the error in $s=\log\sigma$ is $\epsilon_s = 0.025$~dex (6\%), the photometric error is $\epsilon_{\rm  phot} = [\epsilon_r^2 + b\epsilon_i]^{1/2} = 0.022$~dex (5\%), and the intrinsic error in $r$ is $\sigma_{r,{\rm int}} = \sigma_1[1+a^2+b^2]^{1/2}=0.099$~dex (23\%). With these values, the total RMS scatter in $r$ is 25\% and is clearly dominated by the intrinsic scatter. Although we are using a conservative method to calculate $\sigma_{r,{\rm int}}$ (since the additional factors mentioned above tend to reduce the intrinsic scatter), this 25\% scatter in $r$ is still a significant improvement on the 29\% reported by \cite{2012MNRAS.427..245M} (applying the same conservative method to the 6dFGS sample gives 31\% total RMS scatter in $r$). However, the 6dFGS sample is heavily censored, while the SDSS sample is closer to a pure Gaussian distribution. Thus the 29\% reported by \cite{2012MNRAS.427..245M} is the appropriate value of the RMS scatter $\sigma_r$ for 6dFGS to compare with the 25\% scatter found for SDSS. 

The major contribution to this improvement going from 6dFGS to SDSS comes from the smaller errors in the SDSS velocity dispersions. The typical error on the velocity dispersion for the SDSS sample is 6\% (at redshift $z<0.1$) compared to 12\% for the 6dFGS sample. The velocity dispersion error was found to depend primarily on SNR, which in turn depends on factors such as telescope aperture, total exposure time, object flux, and sky flux. However, the {\em intrinsic} scatter of the Fundamental Plane in the $r$ direction is almost the same for these two samples. 

\section{Simultaneous Fundamental Plane and velocity field fits}
\label{FP+Vfits}

In this section we describe a Bayesian forward-modelling approach to simultaneously fit the Fundamental Plane and the velocity field in the space of the observable quantities. This method is similar to the VELMOD method first presented by \cite{1997ApJ...486..629W} to overcome most of the obstacles that faced existing methods at that time, such as POTENT \citep{1994ARA&A..32..371D} and the inverse Tully-Fisher method \citep{1995MNRAS.276.1391N}. The VELMOD method has been applied by several authors, mainly to Tully-Fisher data \citep{1998ApJ...507...64W,2001MNRAS.326.1191B,2015MNRAS.450..317C}.

In the original VELMOD method, the velocity field scaling parameter $\beta$ is treated as a set of discrete values, for each of which the likelihood is maximized. By contrast, we treat $\beta$ and the bulk motion imposed by the external tidal field $\mathbf{V}_{\rm ext}$ as continuous free parameters and we simultaneously seek these velocity field parameters and the Fundamental Plane parameters $a$, $b$, $c$, and $\sigma_r$ using a forward-fitting approach (i.e.\ we predict the observables from the model and fit the data in the observed Fundamental Plane space). Note that $\sigma_r$ is the intrinsic scatter about the Fundamental Plane in the $r$-direction, but differs from the value derived above in fitting the Fundamental Plane {\it only}, because that includes the scatter from peculiar velocities in the total Fundamental Plane scatter, whereas here the peculiar velocities are in principle fitted out by the velocity field model.

\begin{figure}
\begin{center}
\includegraphics[width=\columnwidth]{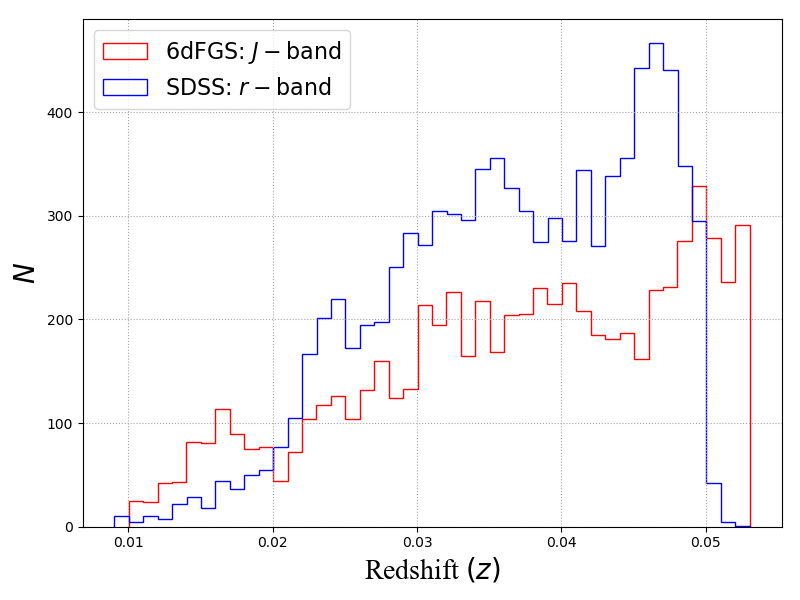}
\caption{The CMB frame redshift distributions for both 6dFGSv $J$-band and SDSS $r$-band peculiar velocity samples.}
\label{cz_6df_sdss_hist}
\end{center}
\end{figure}

The peculiar velocity samples used in this section are subsamples of the Fundamental Plane samples used in section~3. Firstly, with better images and colours now available from the Pan-STARRS1 Surveys \citep{2016arXiv161205560C}, Dark Energy Survey \citep{2018ApJS..235...33D}, the Dark Energy Camera surveys \citep{2015AJ....150..150F}, VISTA Hemisphere Survey \citep{2013Msngr.154...35M}, and SkyMapper Southern Survey \citep{2018PASA...35...10W}, we rejected 1773 galaxies from the 6dFGSv sample that was originally used to fit the 6dF FP \citep{2012MNRAS.427..245M}. For the SDSS sample we limited the redshifts to the same redshift range used for the 6dF survey, i.e.\ $z < 0.055$. Figure~\ref{cz_6df_sdss_hist} shows the CMB frame redshift distribution for the two peculiar velocity samples (7030 6dFGSv and 8864 SDSS galaxies)\footnote{These two catalogs are provided as a supplementary data} that we use in this section. Secondly, we do not use the redshift to convert from angular effective radius to physical effective radius; instead, we do this conversion inside the likelihood using the true distance. 

We calculate the effective redshift for each survey using its limiting and characteristic magnitudes (cf.\ \citealt{1980lssu.book.....P}, section~50). This is a two step method: First, we fit the luminosity function for each sample; Secondly, we use the characteristic depth equation,
\begin{equation}
    D = 10^{0.2(m_0 - M^*)-5} \text{Mpc}
\end{equation}
for each sample, where $m_0$ and $M^*$ are the limiting and characteristic magnitudes. We found the effective redshifts to be $0.033$, $0.036$ and $0.035$ for our 6dFGS, SDSS and combined samples, respectively.

Specifically, we want to compute $P(\theta_e,\sigma_0,I_e,cz,\mathbf{r})$, the joint probability that a galaxy has a redshift $cz$ and Fundamental Plane observables $\theta_e$ (apparent effective radius in angular units), $\sigma_0$ (central velocity dispersion) and $I_e$ (mean surface brightness within the effective radius) at a comoving location $\mathbf{r}$, given the density and velocity model of 2M$++$ \cite{2015MNRAS.450..317C}, the velocity field scaling parameter $\beta$, and the bulk motion due to the external tidal field $\mathbf{V}_{\rm ext}$.

We express the above joint probability as a product of conditional probabilities that can be easily computed, in the form
\begin{equation}
    P(\theta_e,\sigma_0,I_e,cz,\mathbf{r}) = P(\theta_e,\sigma_0,I_e|r) P(cz|\mathbf{r}) P(\mathbf{r}) ~.
\label{Vel_1}
\end{equation}
It is important here to emphasise that $\theta_e$, ${\sigma}_0$, $I_e$, $cz$ and the direction corresponding to location $\mathbf{r}$ are all observables, while the comoving distance $r$ corresponding to location $\mathbf{r}$ is the only non-observable.

The first term on the right-hand side of Equation~\ref{Vel_1} can be expressed in forward, inverse, or 3D Gaussian form. We write the forward Fundamental Plane as:
\begin{multline}
    P(\theta_e,\sigma_0,I_e|r) = \frac{1}{\sqrt{2\pi}\sigma_{\rm FP}} \\ 
    \exp\left[- \frac{[\theta_e-R_e(\sigma,I_e)/d_A(r)]^2}{2\sigma_{\rm FP}^2}\right]
    \label{vel_2}
\end{multline}
where $\log R_e(\sigma_0,I_e) = a \log \sigma_0 + b \log I_e + c$ is the Fundamental Plane relation, $d_A(r)$ is the angular diameter distance corresponding to comoving distance $r$, and $\sigma_{\rm FP}$ in the denominator combines the error in the observed effective radius and the error in the Fundamental Plane forward model, $\sigma_r$ (the RMS scatter in the $r$-direction), in a quadratic form. 

The second term on the right hand side of Equation~\ref{Vel_1}, $P(cz|r)$, couples the Fundamental Plane observables to the velocity model as
\begin{equation}
    P(cz|\mathbf{r}) = \frac{1}{\sqrt{2\pi}\sigma_{v}} \exp\left[-\frac{(cz-[r+u(\mathbf{r})])^2}{2\sigma_v^2}\right]
    \label{vel_3}
\end{equation}
where $cz$ is the observed redshift (taken to have negligible uncertainty) and $r$ and $u(\mathbf{r})$ are the true comoving distance to the galaxy and its model peculiar velocity along the line of sight. The model peculiar velocity is determined by scaling the normalised predicted velocity field $V_{\rm pred}$ (from 2M$++$; \citealt{2015MNRAS.450..317C}) with the $\beta$ parameter and adding the bulk flow due to the external tidal field $\mathbf{V}_{\rm ext}$, so that
\begin{equation}
    u(\mathbf{r}) = \beta V_{\rm  pred}(\mathbf{r}) + \mathbf{V}_{\rm ext} ~.
\end{equation}
Because the published 2M$++$ peculiar velocities are already multiplied by a fiducial value of $\beta$, and the external dipole $\mathbf{V}_{\rm ext}$ has already been added, we first subtract the fiducial value of $\mathbf{V}_{\rm ext}$ ($v_x = 89$, $v_y = -131$, $v_z =17$; \citealt{2015MNRAS.450..317C}) and then divide by the fiducial value of
$\beta$ ($\beta = 0.43$; \citealt{2015MNRAS.450..317C}). Note that here we are removing the values of $\beta$ and $\mathbf{V}_{\rm ext}$ that were added at later stages and not during the reconstruction process itself. The quantity $\sigma_v$ in the denominator of Equation~\ref{vel_3} parametrises the residual variation in the observed non-linear velocity field relative to the predicted linear velocity field; for consistency with \cite{2015MNRAS.450..317C} we adopted $\sigma_v = 150$~km~s$^{-1}$.

The third term on the right hand side of Equation~\ref{Vel_1}, $P(\mathbf{r})$, is the probability of observing a galaxy at comoving location $\mathbf{r}$ (with comoving distance $r$), and is given by 
\begin{equation}
    P(\mathbf{r}) \propto r^2 (1+\delta_g(\mathbf{r}))   
    \label{vel_4}
\end{equation}
where the $r^2$ term accounts for homogeneous Malmquist bias, the $1+\delta_g(\mathbf{r})$ term accounts for inhomogeneous Malmquist bias, and $\delta_g(\mathbf{r})$ is the number density at location $\mathbf{r}$ in the 2M$++$ density field model \citep{2015MNRAS.450..317C}. 

Substituting Equations~\ref{vel_2}, \ref{vel_3} and~\ref{vel_4} into Equation~\ref{Vel_1} we obtain the final expression for the joint probability:
\begin{multline}
P(\theta_e,\sigma_0,I_e,cz,\mathbf{r}) = \\
            \frac{1}{\sqrt{2\pi}\sigma_{\rm FP}}
            \exp\left[- \frac{[\theta_e - R_e(\sigma_0,I_e)/d_A(r)]^2}
            {2\sigma_{\rm FP}^2}\right] \\ 
            \frac{1}{\sqrt{2\pi}\sigma_{v}}
            \exp\left[-\frac{(cz-[r+u(\mathbf{r})])^2}{2\sigma_v^2}\right] r^2 (1+\delta_g(\mathbf{r})) ~.
            \label{vel_all}
\end{multline}
We can then write the likelihood function as
\begin{equation}
    \Lagr = \prod_{n=1}^{N_g} P_n(\theta_e,\sigma_0,I_e,cz,\mathbf{r})^{1/S_n}.
\end{equation}
although in practice we use the log-likelihood given by
\begin{multline}
   \ln \Lagr = -\frac{1}{2} \sum_{n=1}^{N_g} \frac{1}{S_n} \Bigg[\frac{[\theta_e - R_e(\sigma_0,I_e)/d_A(r)]^2}{\sigma_{FP}^2} + \ln(\sigma_{FP}^2)\\
   + \frac{(cz-[r+u(\mathbf{r})])^2}{\sigma_v^2} + \ln(\sigma_v^2) - 2 \ln(r^2(1+\delta_g(\mathbf{r}))) \Bigg]_n
   \label{vel_likelihood}
\end{multline}
where $S_n$ is the selection probability as in the 3D Gaussian fitting above.

We want to solve Equation~\ref{vel_likelihood} from a Bayesian point of view---in other words, we want to determine the posterior probability function specified by the Fundamental Plane parameters ($a$, $b$, $c$, and $\sigma_r$) and the velocity field parameters ($\beta$ and $\mathbf{V}_{\rm ext}$) that is consistent with the set of Fundamental Plane observables ($\theta_e$, ${\sigma}_0$, $I_e$, and $cz$) by marginalizing over the nuisance parameters (the unknown distances to each galaxy $r_n$). Using Markov chain Monte Carlo (MCMC) allows us to do this in a single step.

With the likelihood function in hand, the remaining information required for the MCMC is the prior probability function $P(a,b,c,\sigma,\beta,\mathbf{V}_{\rm ext},r_n)$ that captures all previous knowledge about the parameters. We employ a simple uninformative uniform prior that requires $a,b,c$ and $\sigma$ to be between 10 and $-10$, distances to be in the range 10~km~s$^{-1} < r < 20000$~km~s$^{-1}$, the velocity scaling parameter to be in the range $0 < \beta < 2$, and no prior knowledge on the external tidal velocity amplitude or direction.

The parameter $\beta$ depends on the clustering of the galaxies in the selected sample, which in turn depends on their mass and morphology and the waveband in which they are observed \citep{2007PhDT.........3W}. Thus to be consistent in the use of this parameter and to facilitate comparisons between samples, we need to apply a wavelength- and luminosity-dependent correction to calibrate the 6dFGSv and SDSS samples to the 2M$++$ survey. This correction is significantly larger in near-infrared than in optical bands \citep{2001Natur.410..169P,2001MNRAS.328...64N,2007PhDT.........3W}.

For each galaxy in the $J$-band 6dFGSv sample we measured the absolute magnitude in the $K_s$-band using the same method as the 2M$++$ catalogue \citep{2011MNRAS.416.2840L}. The apparent magnitude is measured as the isophotal magnitude at 20~mag~arcsec$^{-2}$. The dust extinction correction, evolution correction, and $K$-correction were also as applied by \cite{2011MNRAS.416.2840L}. We found that the mean absolute magnitude for the 6dFGS peculiar velocity sample is brighter than the 2M$++$ characteristic absolute magnitude by 0.23~mag, corresponding to a factor $\Bar{L}/L_*=1.24$. Using the relation between galaxy bias and luminosity in the near-infrared given by \cite{2007PhDT.........3W}, $b/b_*=0.73+0.24L/L_*$, this implies that the 6dFGS sample is 1.03 more clustered than the 2M$++$ sample.

For the SDSS sample, we used the New York University Value-Added Galaxy Catalog (NYU-VAGC; \citealt{2005AJ....129.2562B}) to find the counterpart for each galaxy in the SDSS $r$-band sample. We then used the same method as above to calculate the $K_s$-band absolute magnitude to compare to the 2M$++$ catalogue. In contrast to the 6dFGS sample, the SDSS sample was found to be dimmer than the 2M$++$ sample by 0.463~mag, implying $\Bar{L}/L_* = 0.65$. Applying a less steep relation for optical bands, $b/b_*=0.85+0.15L/L_*$, given by \cite{2001MNRAS.328...64N}, suggests that the SDSS pv sample used here is 0.95 less clustered than the 2M$++$ sample.

The above $b/b*$ values were used to convert 2M$++$ $\delta_g^*$ to 6dFGS $\delta_g^{\text{6dF}}$ and SDSS $\delta_g^{\text{SDSS}}$ which are then used in place of $\delta_g$ in equation (\ref{vel_likelihood}). This correction makes unnoticeable change to our results.    

Using this approach we have been able to constrain the Fundamental Plane parameters $a$, $b$, $c$ and $\sigma_r$ as well as the velocity field parameters $\beta$, ${V}_{\rm x}$, ${V}_{\rm y}$, and ${V}_{\rm z}$ from the 6dFGSv and SDSS samples. The fitted values of these parameters for the 6dFGSv sample, and their estimated errors, are given in Table \ref{forward_6dF+SDSS}.

\begin{table}
\begin{center}
\caption[]{Fundamental Plane and velocity field parameters, and their 68\% confidence intervals, for the 6dFGSv $J$-band and SDSS $r$-band samples.}
\begin{tabular} {lcc}
\hline\hline
 Parameter    & 6dFGSv            & SDSS \\
\hline
$N_g$         & 7030               & 8864 \\
$a$           & ~~~$0.916\pm0.028$ & ~~~$0.844\pm0.015$\\
$b$           &   $-0.895\pm0.010$ &   $-0.930\pm0.009$\\
$c$           &   $-0.846\pm0.037$ &   $-1.200\pm0.035$\\
$\sigma_r$    & ~~~$0.095\pm0.006$ & ~~~$0.089\pm0.003$\\
$\beta$       & ~~~$0.372\pm0.042$ & ~~~$0.314\pm0.039$\\
$V_{\rm x}$ & ~~~$91\pm9$~km~s$^{-1}$ & ~~~$98\pm9$~km~s$^{-1}$\\
$V_{\rm y}$ & ~~~$-127\pm14$~km~s$^{-1}$ & ~~~$-148\pm13$~km~s$^{-1}$\\
$V_{\rm z}$ & ~~~$-4\pm8$~km~s$^{-1}$ & ~~~$12\pm9$~km~s$^{-1}$\\
$||V||$ & ~~~$156\pm13$~km~s$^{-1}$ & ~~~$178\pm12$~km~s$^{-1}$\\
l & ~~~$306\pm4$~degree & ~~~$303\pm4$~degree\\
b & ~~~$-2\pm3$~degree & ~~~$4\pm3$~degree\\
\hline
\end{tabular}
\label{forward_6dF+SDSS}
\end{center}
\end{table}

\begin{figure*}
\begin{center}
\includegraphics[width=\textwidth]{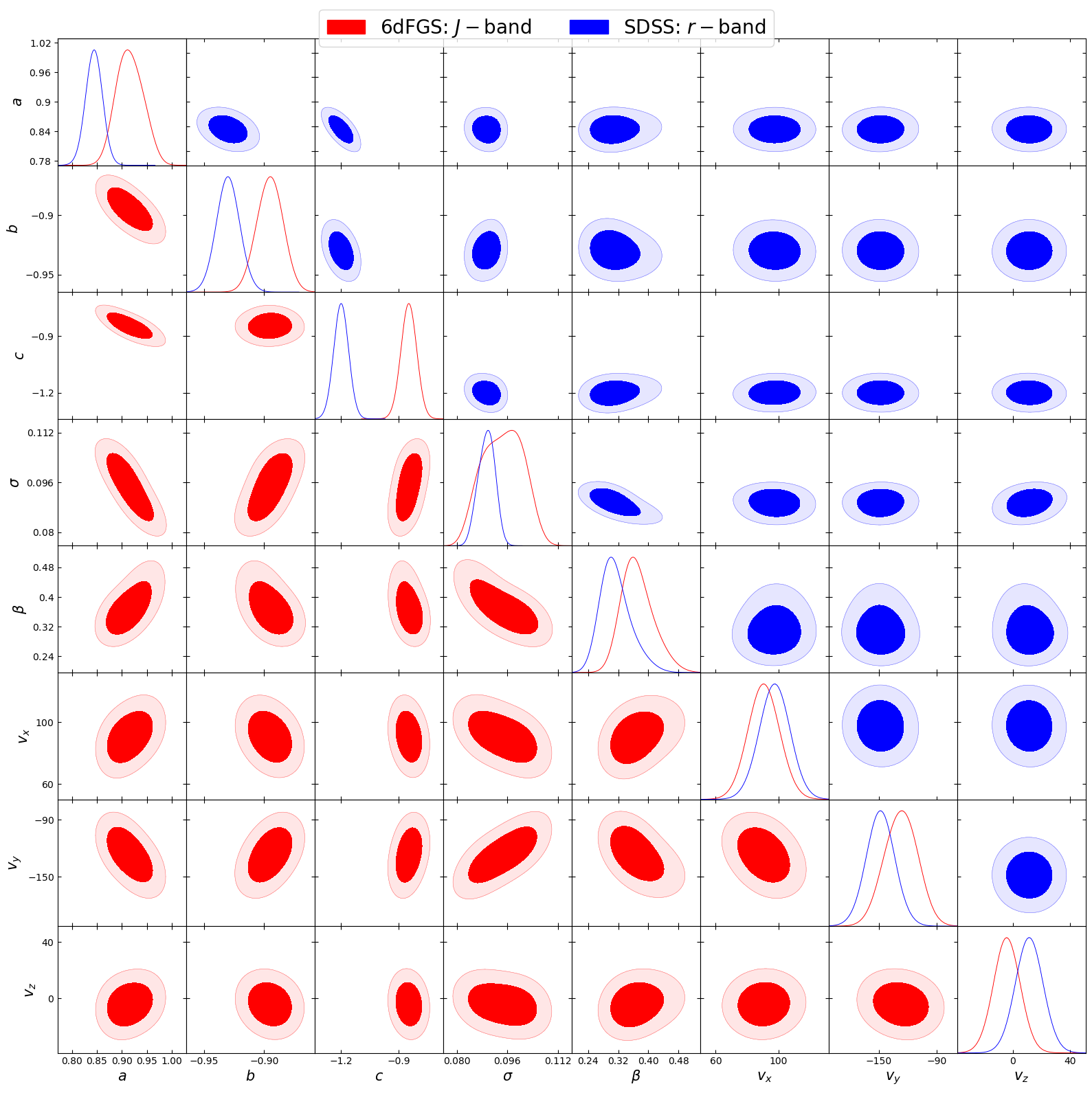}
\caption{The pairwise joint constraints on $a$, $b$, $c$, $\sigma_r$, $\beta$, $V_{\rm x}$, $V_{\rm y}$, and $V_{\rm z}$ from fitting the 6dFGSv $J$-band (red) and SDSS $r$-band (blue), Fundamental Plane data using the 2M$++$ model for the density and peculiar velocity fields. The dark and light blue shadings show, respectively, the 68\% and 95\% confidence regions.}
\label{corner_150}
\end{center}
\end{figure*}

Figure~\ref{corner_150} shows the pairwise joint constraints on $a$, $b$, $c$, $\sigma_r$, $\beta$, ${V}_{\rm x}$, ${V}_{\rm y}$, and ${V}_{\rm z}$ from fitting the 6dFGSv $J$-band (red) and SDSS $r$-band (blue) Fundamental Plane data using the 2M$++$ model for the density and peculiar velocity fields. As expected there are weak correlations between the slopes and intercept of the Fundamental Plane ($a$, $b$, $c$) and between the intrinsic scatter about the Fundamental Plane ($\sigma_r$) and the velocity field parameters ($\beta$ and $\bf{V}_{\rm ext}$), in the sense that larger intrinsic scatter about the Fundamental Plane corresponds to smaller peculiar velocities (smaller $\beta$) and a lower amplitude of the bulk velocity due to the tidal field (lower $\bf{V}_{\rm ext}$). Similarly, larger peculiar velocities (larger $\beta$) weakly correlate with lower amplitude of the bulk velocity due to the tidal field (lower $\bf{V}_{\rm ext}$).

\begin{figure*}
\begin{center}
\includegraphics[width=\textwidth]{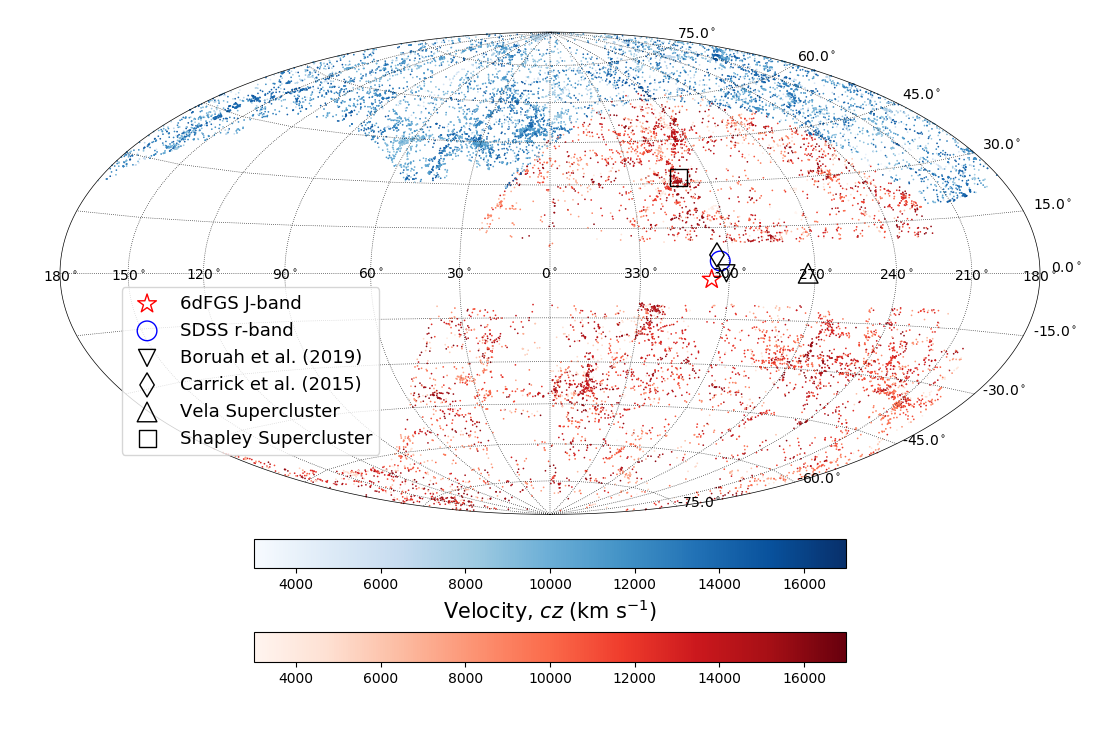}
\vspace*{-13mm}
\caption{The distribution of the 15894 elliptical galaxies in 6dFGSv (Reds) and SDSS (Blues) pv samples colour-coded by their redshift shown in an Aitoff projection. The direction of the external bulk flow $\bf{V}_{\rm ext}$ is shown as a red star for 6dFGSv sample and as a blue circle for SDSS sample. We also plot the direction of external bulk flow derived by \protect\cite{2015MNRAS.450..317C} as a thin diamond and \protect\cite{2019arXiv191209383B} as triangle down as well as the position of the two most prominent superclusters in the nearby universe, Vela supercluster as black triangle up and Shapley Supercluster as a square.}
\label{6df_bulk_flow_redshift}
\end{center}
\end{figure*}

Figure \ref{6df_bulk_flow_redshift} shows the distribution of 6dFGSv (colour-coded as reds) and SDSS (colour-coded as blues) peculiar velocity samples in an Aitoff projection. The direction of the 6dFGSv residual bulk flow is shown as a red-star, while the direction of the SDSS residual bulk flow is presented by the blue circle. Although 6dFGSv and SDSS sample cover different volume of space, they agree on the direction of the residual bulk flow. There is also a good agreement for both amplitude and direction of the external bulk flow with \cite{2015MNRAS.450..317C} who reported $||V|| = 159\pm23$ km s$^{-1}$, $l = 304\pm11$ deg, and $b=6\pm13$ deg as well as \cite{2019arXiv191209383B} who reported $||V|| = 171\pm11$ km s$^{-1}$, $l = 301\pm4$ deg, and $b=0\pm3$ deg. We also highlighted the position of two of the most prominent structures in the local Universe, Vela supercluster (triangle; \citealt{2017MNRAS.466L..29K}) and Shapley Supercluster (diamond). The 6dFGSv and SDSS residual velocity directions are fully consistent and lie midway between the two biggest superclusters (Shapley and Vela) known to lie at the edge of the survey volume.

A comparison of the fitted velocity field parameters from both 6dFGSv and SDSS with previous literature results is given in Figure~\ref{rectangle_6df_sdss_long}. It shows the marginalized constraint contours for the peculiar velocity scaling parameter $\beta$ and the three components of external velocity $\bf{V}_{\rm ext}$ from both 6dFGSv and SDSS samples against the results from three recent studies by \cite{2011MNRAS.413.2906D}, \cite{2012MNRAS.423.3430B}, and \cite{2015MNRAS.450..317C}. Our results give similar values of $\beta$ and $\bf{V}_{\rm ext}$ as these studies. It is remarkable that our fits for $\beta$ and $\bf{V}_{\rm ext}$ from the 6dFGSv $J$-band and SDSS $r$-band samples agree with each other given the different wavebands, sample selection algorithms, and the nearly disjoint volumes probed by the two samples. Our results are in 1-2$\sigma$ agreement with the inverse Tully-Fisher analysis by \cite{2011MNRAS.413.2906D}, the RSD analysis by \cite{2012MNRAS.423.3430B}, and the forward likelihood Tully-Fisher analysis by \cite{2015MNRAS.450..317C}, despite using different galaxy types, volumes of space, distance indicators and methods of analysis.

\begin{figure*}
\begin{center}
\includegraphics[width=\textwidth]{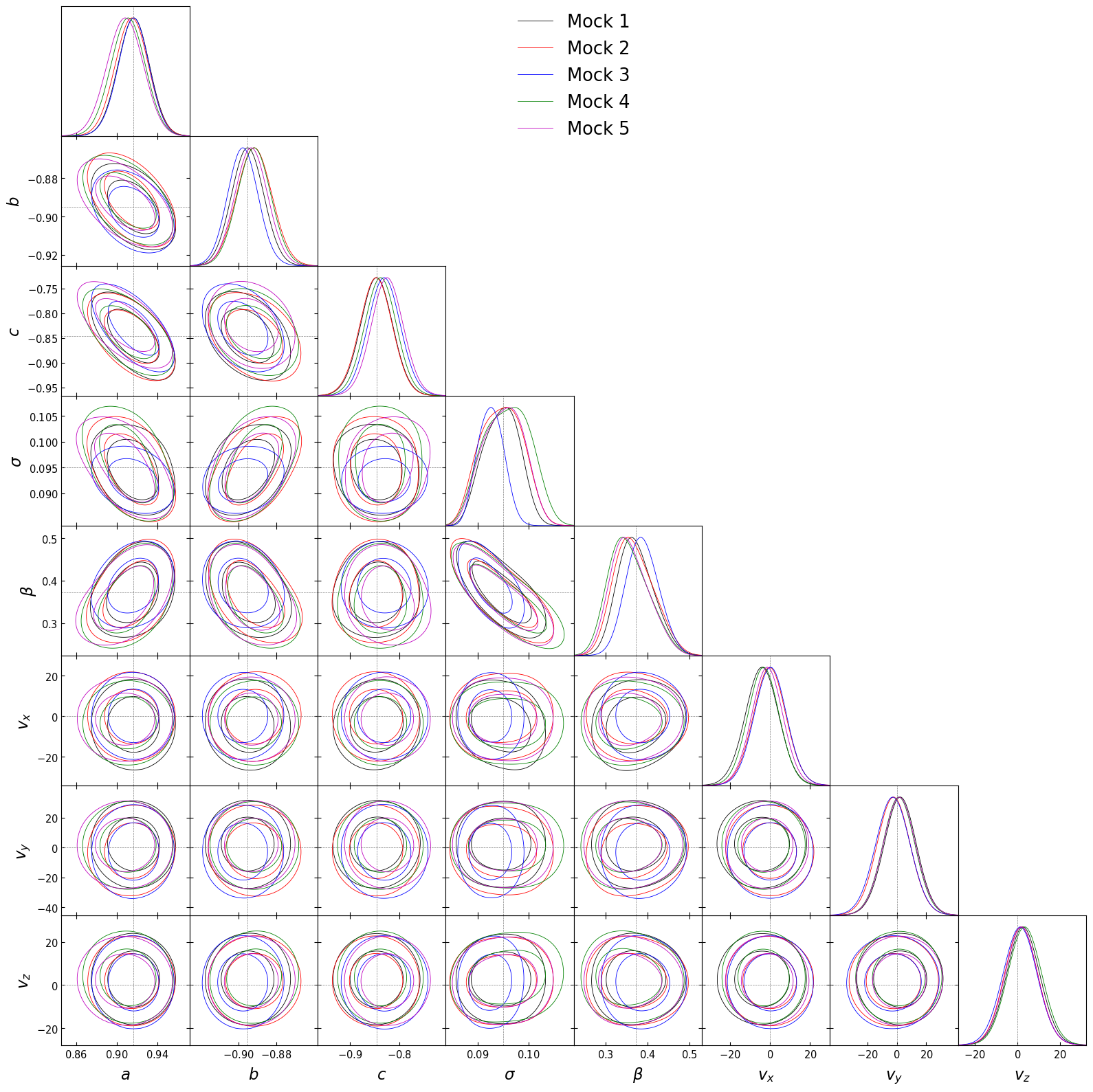}
\caption{The pairwise joint constraints on $a$, $b$, $c$, $\sigma_r$, $\beta$, $V_{\rm x}$, $V_{\rm y}$, and $V_{\rm z}$ from fitting the 6dFGSv $J$-band mocks. The inner and outer contours  show, respectively, the 68\% and 95\% confidence regions. The different colours indicate different mocks. The input parameters are shown as a dotted lines.}
\label{corner_6df_mocks_n}
\end{center}
\end{figure*}

As in Section~3, we validated our method using mocks. We generated five mocks, following the approach of \citep{2012MNRAS.427..245M}, to closely match the observed peculiar velocity sample of 6dFGS. To do this, we refitted the 6dF peculiar velocity sample using the 3D Gaussian method used in Section~3. The fitted parameters were then used to generate the new mocks. Peculiar velocities were assigned the 2M$++$ predicted velocity with the fitted external bulk flow added to it. We clone each mock galaxy into the opposite hemisphere in order to avoid any systematic dipole that might affect the recovered mock parameters; the clone galaxy is identical except that the sign of the Declination is flipped and the Right Ascension is rotated 180 degrees. For each cloned galaxy the same  predicted velocity as the original one was assigned from 2M++ model but with flipped sign. Thus, predicted velocities comes only from actual structures in the south and we cancel its effect by using the same velocity but with flipped sign in the north. This should lead to a zero external bulk flow in our fitted parameters without affecting other quantities such as the Fundamental Plane parameters and $\beta$. This also should lead to a better constraints on all parameters given the double size of the sample.  Figure~\ref{corner_6df_mocks_n} shows the pairwise joint constraints on $a$, $b$, $c$, $\sigma_r$, $\beta$, $V_{\rm x}$, $V_{\rm y}$, and $V_{\rm z}$; input parameter values are shown by dotted lines. The figure demonstrates that our method of simultaneously fitting Fundamental Plane and velocity field parameters recovers the input parameters with high precision.

\section{Discussion}

The determination of the 3D Gaussian Fundamental Plane using a sample of galaxies from SDSS $r$-band shows a significant improvement, in the sense of reduced scatter, compared to the analysis using the 6dFGSv sample (Section~\ref{FPfits} and Table~\ref{3D_gaussian_6df_sdss}). The main improvement comes from the difference in the typical uncertainty in the velocity dispersion measurements between the two samples. Both Fundamental Plane relations have almost the same intrinsic scatter (23\%), but the SDSS sample has a typical velocity dispersion error of 6\% compared to 12\% for the 6dFGSv sample. This leads to a corresponding improvement from 29\% for 6dFGSv to 25\% for SDSS in the total scatter of the Fundamental Plane in the $r$ direction, $\sigma_r$, which is proportional to the true distance error $\sigma_d$.  

The Taipan galaxy survey (Taipan; \citealt{2017PASA...34...47D}) will improve on 6dFGSv and be closer to SDSS in this respect by using repeat observations to build up the spectral SNR, which is inversely proportional to the uncertainty on the velocity dispersion measurements. Future surveys like the Dark Energy Spectroscopic Instrument (DESI; \citealt{2016arXiv161100036D}) will be able to improve further by observing each galaxy with several fibres (made possible for DESI by the density of fibres in the focal plane). The obvious way to locate the fibres will be along the major and minor axes as much as possible. That will be rotation velocity for spirals to be used for Tully-Fisher and velocity dispersion for elliptical galaxies to be used for Fundamental Plane. The resultant velocity dispersion measurement will be better than most previous Fundamental Plane surveys, because it will be able to measure the velocity dispersion at multiple radii, instead of just the central value. This will improve the Fundamental Plane scatter by $\sim 24\%$ \citep{2017ApJ...843...74O} and consequently distances can be inferred more accurately.

The Fundamental Plane parameters are highly dependent on the waveband and sample properties, which makes it hard to compare results obtained from different samples. In contrast, the $\beta$ parameter and externally-induced bulk flow $\bf{V}_{\rm ext}$ in principle only weakly depend on the bias parameters $b$ of the samples from which they are derived ($\delta_g(\mathbf{r}$) term in equation \ref{vel_likelihood}). Comparing $\beta$ from different samples thus requires additional information about their bias values. Selecting a galaxy sample in the $K_s$-band (as is the case for the 6dFGSv sample) will be biased towards larger, brighter, more clustered galaxies, whereas selecting in the $r$-band (as is the case for SDSS) will lead to a less biased combination of field and cluster galaxies. In this paper we have corrected for this problem using empirical relations for the dependence of the linear bias factor on waveband and luminosity (\citealt{2007PhDT.........3W,2001MNRAS.328...64N}).  

\begin{figure*}
\begin{center}
\includegraphics[width=\textwidth]{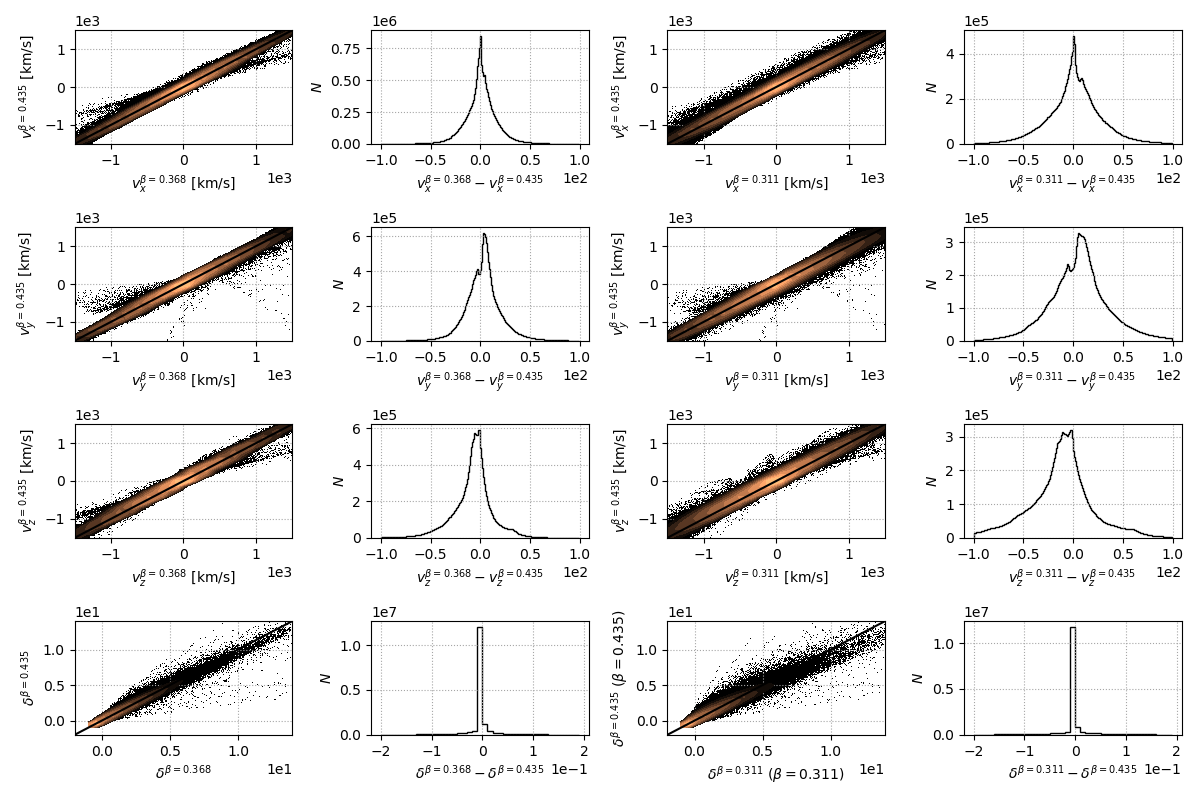}
\caption{Comparison of different density and velocity field models using different values of $\beta$. The top three rows show the comparison between velocity fields in Galactic Cartesian comoving coordinates X, Y, Z. The bottom row presents the comparison between different density field models. The largest offset from zero was found to be 10 times smaller than the error in the velocity field $\sigma_v$. The largest standard deviation was also found to be 4 times smaller than $\sigma_v$. The offset in the density field model was of order $10^{-4}$.}
\label{density_velocity_6df_sdss}
\end{center}
\end{figure*}

We checked various other effects that could bias our measurement of $\beta$. One potential issue is using velocity and density field models that were reconstructed with a fiducial value of $\beta$. We checked this effect by comparing the predicted velocities and densities reconstructed with different values of $\beta$. We were privately provided with the full set of 2M$++$ \citep{2015MNRAS.450..317C} models, which includes density and velocity fields reconstructed with values of $\beta$ ranging from 0.01 to 0.86. We used three values of $\beta$, 0.435 (the published value for 2M$++$), 0.368 (the nearest value to our 6dF fitted value), and 0.311 (the nearest value to our SDSS fitted value). Figure~\ref{density_velocity_6df_sdss} shows the comparison between predicted velocities in Galactic Cartesian comoving coordinates X,Y,Z (the top three rows) and densities (bottom row) reconstructed with different values of $\beta$. The first two columns show the comparison using $\beta=0.368$ and $\beta=0.435$; out of $\sim$17 million data points, only a few deviate from the one-to-one line. We quantitatively measured any systematic using the mean and the standard deviation: for $v_x$, $v_y$, and $v_z$, the mean and standard deviation (in km\,s$^{-1}$) are 0.5 and 19, 3 and 22, and $-$8 and 23 respectively. The same was comparison was performed using $\beta=0.311$ and $\beta=0.435$; in that cases the mean and standard deviation (in km\,s$^{-1}$) are 1 and 32, 7 and 37, and $-$15 and 40 respectively. The largest offset from the mean was found to be in $v_z$, but it is 10 times smaller than the error we used for each individual predicted velocity value ($\sigma_v=150$\,km\,s$^{-1}$). On the other hand, the largest standard deviation was also found to be in $v_z$, and is 4 times smaller than $\sigma_v$. The offset and standard deviation for the density field was much smaller, of order $10^{-4}$. Thus we concluded it was entirely adequate to only use the published 2M$++$ density and velocity fields through out our analysis.

\begin{figure}
\begin{center}
\includegraphics[width=\columnwidth]{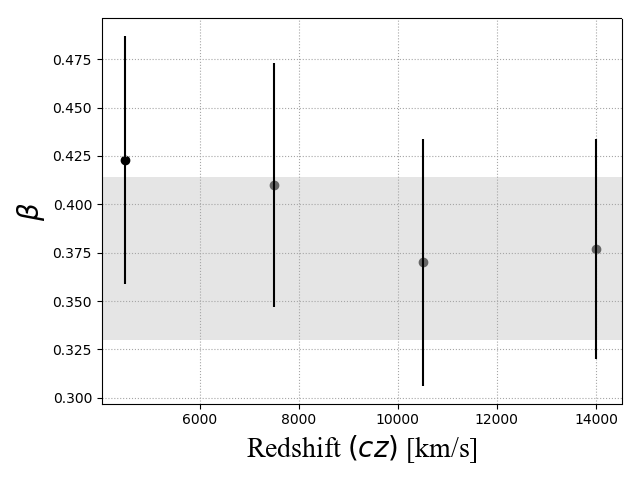}
\caption{The fitted 6dF $\beta$ values in independent shells of redshift. The number of galaxies at each shell, from low to high redshift, is 651, 1116, 2039, 3224. Each value of $\beta$ is centered at the mean redshift of the shell. The grey shade presents our overall $\beta$ value for the 6dF sample.}
\label{betavsredshift}
\end{center}
\end{figure}

The second effect that could bias our value of $\beta$ is a dependence on redshift. We checked this effect by dividing our 6dFGS sample into four independent shells in redshift. Figure~\ref{betavsredshift} shows the fitted values of $\beta$ for each redshift shell; note that the larger peculiar velocity errors for individual galaxies in the higher redshift shells are approximately compensated by the increased number of galaxies in those shells. There is a weak trend of decreasing $\beta$ with redshift, but all shells are consistent with each other within the errors. Our overall value for $\beta$ (with $\pm$1$\sigma$ error range shown in grey) agrees more with the high redshift values---but again all values agree within the errors.

\begin{figure*}
\begin{center}
\includegraphics[width=\textwidth]{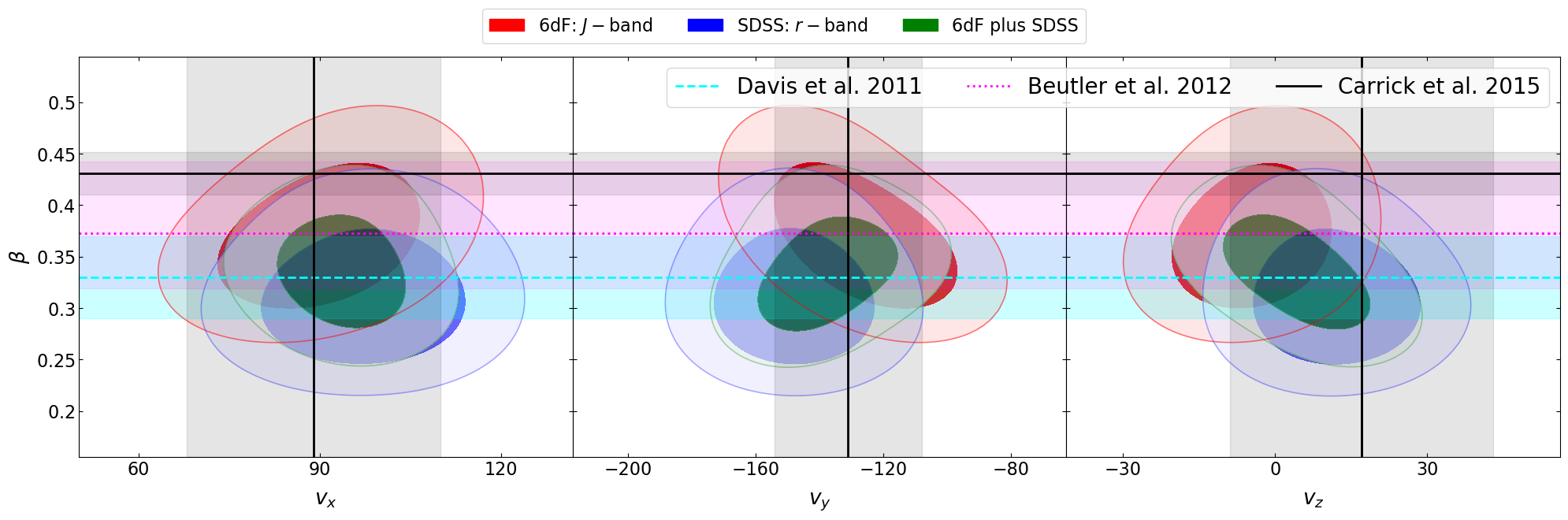}
\caption{Constraints on $\beta$ and $\bf{V}_{\rm ext}$ from 6dFGSv (red), SDSS (blue), and their combination (green) compared to literature results. The contours show 68\% and 95\% confidence ranges. The dashed-cyan, dotted-magenta and solid-grey lines (and correspondingly-coloured bands) show the measurements (and $1\sigma$ confidence intervals) reported respectively by \protect\cite{2011MNRAS.413.2906D}, \protect\cite{2012MNRAS.423.3430B}, and \protect\cite{2015MNRAS.450..317C}.}
\label{rectangle_6df_sdss_long}
\end{center}
\end{figure*}

With the assumption that the corrections we have applied for luminosity effects using the empirical relations hold, we combine our two samples from 6dFGSv and SDSS in Figure~\ref{rectangle_6df_sdss_long}. The figure shows the 68\% and 95\% confidence regions for the $\beta$ parameter and the residual bulk flow $\bf{V}_{\rm ext}$. The 6dFGSv (red) and SDSS (blue) results overlap in the $1\sigma$ region of the parameter space. The confidence regions for the combined 6dFGSv+SDSS sample (green) show the joint constraints, with the single-parameter best fits and 68\% confidence intervals being $\beta=0.341\pm0.024$, $V_{\rm x}=94\pm10$~km~s$^{-1}$, $V_{\rm y}=-138\pm12$~km~s$^{-1}$, and $V_{\rm z}=4\pm12$~km~s$^{-1}$. We did not refit for the Fundamental Plane parameters, instead taking the respective values in Table~\ref{forward_6dF+SDSS} as fixed for 6dFGS and SDSS, and only fitting for $\beta$ and $V_{\text{ext}}$.

One of the key reasons for measuring the $\beta$ parameter is to constrain the growth rate of cosmic structure, $f\sigma_8$. Combining $\beta$ with $\sigma_8,g=0.99\pm0.04$ from the 2M$++$ survey \citep{2015MNRAS.450..317C} suggests that $f\sigma_8 = 0.338\pm0.027$ (8\% uncertainty). The parameter $\sigma_8,g$ is the RMS fluctuations in galaxy number within spherical volumes of radius 8~$h^{-1}$~Mpc. This value was independently calculated by \cite{2015MNRAS.450..317C} using redshift data only, following the method of counts in cells proposed by \cite{1990MNRAS.247P..10E}. This is consistent with the results of \cite{2012MNRAS.424..472B}, who reported $f\sigma_8=0.31\pm0.09$ by combining their $\beta$ parameter with a value of $\sigma_{8,g}$ for the 2MASS redshift survey obtained by \cite{2007PhDT.........3W}. They also calculated $f\sigma_8=0.31\pm0.06$ using the $\beta$ value from \cite{2011MNRAS.413.2906D}. It is important to note that, both \cite{2015MNRAS.450..317C} and \cite{2019arXiv191209383B} used correction for the nonlinear evolution at late times proposed by \cite{2010JCAP...02..021J}. Using the same correction for our 6dF+SDSS sample gives a value of $f\sigma_8 = 0.311\pm0.027$ which is in even a better agreement with \cite{2011MNRAS.413.2906D,2012MNRAS.424..472B}. Throughout this work, we quote values without that correction as most of the other analyses that we are comparing with do not use it as well.

\begin{figure*}
\begin{center}
\includegraphics[scale=0.38]{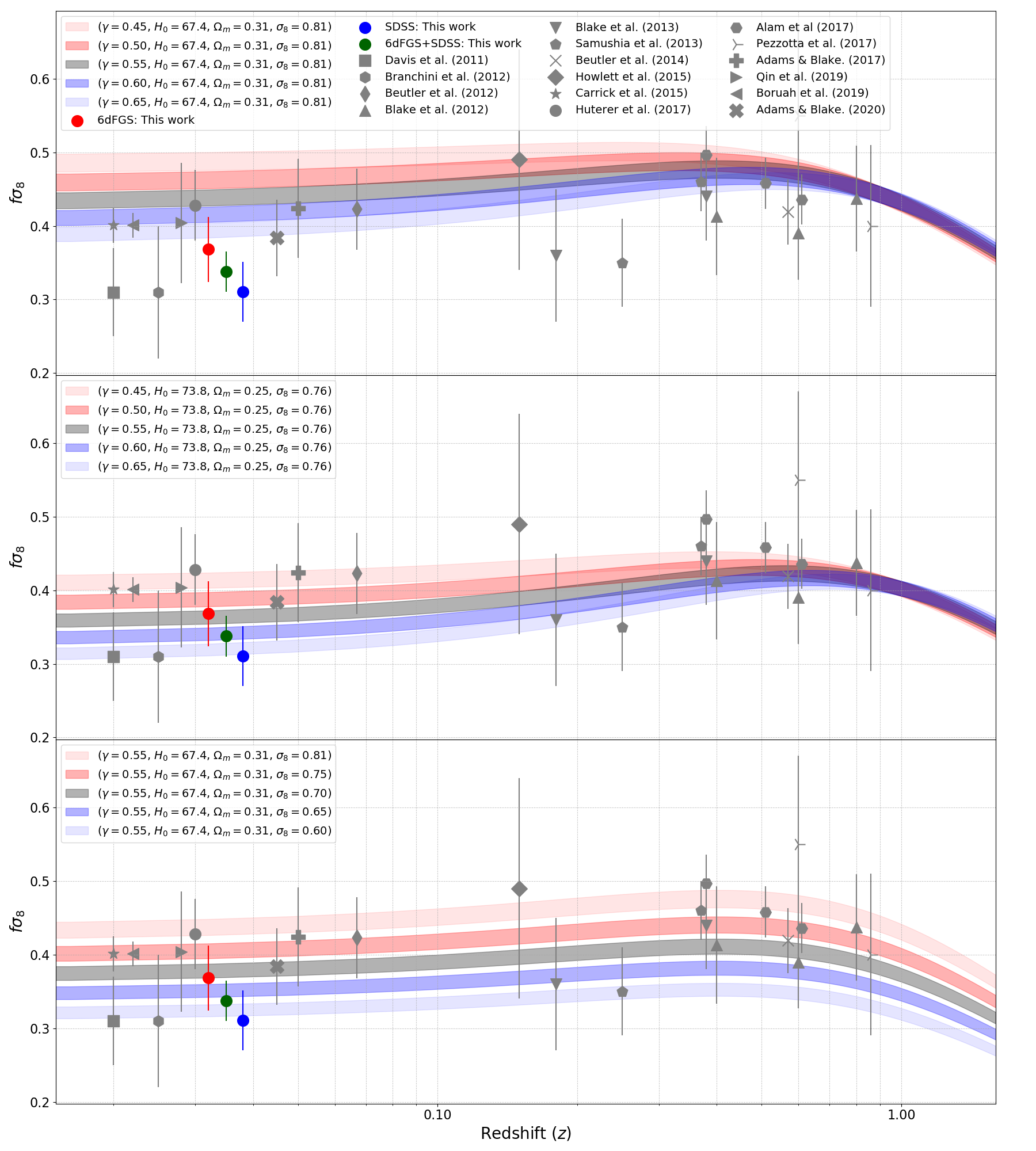}
\caption{Constraints on the growth rate of structure using different galaxy surveys at various effective redshifts. The different data points in all subplots are: 6dFGSv peculiar velocities (red dot, offset by $z = -0.003$ for clarity; This paper), SDSS peculiar velocities (blue dot, offset by $z = +0.003$ for clarity; this paper), 6dFGSv+SDSS peculiar velocities (green dot; this paper), \protect\citealt{2011MNRAS.413.2906D} (square), \protect\citealt{2012MNRAS.424..472B} (hexagon), \protect\citealt{2012MNRAS.423.3430B} (diamond), \protect\citealt{2012MNRAS.425..405B} (triangle-up), \protect\citealt{2013MNRAS.436.3089B} (triangle-down), \protect\citealt{2013MNRAS.429.1514S} (pentagon), \protect\citealt{2014MNRAS.443.1065B} (cross), \protect\citealt{2015MNRAS.449..848H} (diamond), \protect\citealt{2015MNRAS.450..317C} (star), \protect\citealt{2017JCAP...05..015H} (circle, offset by $z = +0.01$ for clarity), \protect\citealt{2017A&A...604A..33P} (star),
\protect\citealt{2017MNRAS.470.2617A} (hexagon2),
\protect\citealt{2017MNRAS.471..839A} (filled plus),
\protect\citealt{2019MNRAS.487.5235Q} (triangle right),  \protect\citealt{2019arXiv191209383B} (triangle left), and
\protect\citealt{2020MNRAS.494.3275A} (filled cross). Top-panel: different bands are using Planck parameters \protect\citep{2018arXiv180706209P} and different $\gamma$. Mid-panel: different bands are using the 6dFGSv cosmological parameters as given by \protect\citep{2012MNRAS.423.3430B} which assumes $H_0=73.8\pm2.4$ and different $\gamma$. Bottom-panel: different bands are using $\gamma=0.55$, $H_0=67.4$, $\Omega_m=0.31$, and different $\sigma_8$.}
\label{testing_gr}
\end{center}
\end{figure*}

Figure \ref{testing_gr} shows a comparison of our three measurements of $f\sigma_8$ from this paper (from 6dFGSv, SDSS and 6dFGSv+SDSS) and several other measurements obtained at higher effective redshifts from a variety of galaxy redshift surveys. As shown, there is a good agreement between measurements at low redshift ($z<0.05$), which is an exclusive scale for peculiar velocity analyses, from this paper, \cite{2011MNRAS.413.2906D}, and  \cite{2012MNRAS.424..472B}. However, some studies at this scale (such as \cite{2017JCAP...05..015H}, suggest a higher value of $f\sigma_8$ which still agrees within the uncertainty with our 6dF value. 

We also show measurements of the growth rate of structure derived from redshift space distortions in various redshift surveys: 6dFGS \citep{2012MNRAS.423.3430B}, WiggleZ \citep{2012MNRAS.425..405B}, GAMA \citep{2013MNRAS.436.3089B}, SDSS-III \citep{2013MNRAS.429.1514S,2014MNRAS.443.1065B}, SDSS MGS \citep{2015MNRAS.449..848H}, BOSS DR12 \citep{2017MNRAS.470.2617A}, and VIPERS \citep{2017A&A...604A..33P}. The figure also includes several coloured bands each of which corresponds to $f\sigma_8$ as a function of redshift as obtained from different theories of gravity or different cosmological parameters. Top panel: the grey band shows a standard $\Lambda$CDM model with $\gamma=0.55$ for General Relativity (GR) and Planck parameters \protect\citep{2018arXiv180706209P}. Deviations from GR are illustrated by the same model with different values of $\gamma$. Mid-panel: the grey band adopts $\gamma=0.55$ and the 6dFGS cosmological parameters from \cite{2012MNRAS.423.3430B} which assumes a prior Hubble constant $H_0=73.8\pm2.4$ from \cite{2011ApJ...730..119R} and $\gamma=0.55$. Again, Deviations from GR are illustrated by the same model with different values of $\gamma$. Bottom-panel: the bands use $\gamma=0.55$, $\Omega_m =0.31$, and different $\sigma_8$.

Assuming values of $\Omega_m$ and $\gamma$ one can estimate a model-dependent value for $f$. Combining the resultant value of $f$ with our measurement of $f\sigma_8$ we can constrain $\sigma_8$ at low redshift. Adopting $\Omega_m = 0.315\pm0.007$ from the \cite{2018arXiv180706209P} and $\gamma=6/11$ for GR from \cite{2007APh....28..481L} gives $f=0.55$. Combining this value with our $\beta=f/b=0.341\pm0.024$ parameter gives a bias factor of $b=1.63\pm0.12$. Using the 2M$++$ value of $\sigma_{8,g}$ from \cite{2015MNRAS.450..317C} suggests a value of $\sigma_8$($z=0.035$) = $\sigma_{8,g}/b = 0.612\pm0.051$ and hence $\sigma_8$($z=0$) = $0.622\pm0.052$. Therefore, at redshift zero, $S_8$ ($z=0$) = $\sigma_8\sqrt{\Omega_m/0.3} = 0.637\pm0.054$. Using the same assumptions except for $\gamma=11/16$ for DGP from \citealt{2007APh....28..481L} gives $S_8$ ($z=0$) = $\sigma_8\sqrt{\Omega_m/0.3} = 0.741\pm0.062$. 

\begin{figure}
\begin{center}
\includegraphics[width=\columnwidth]{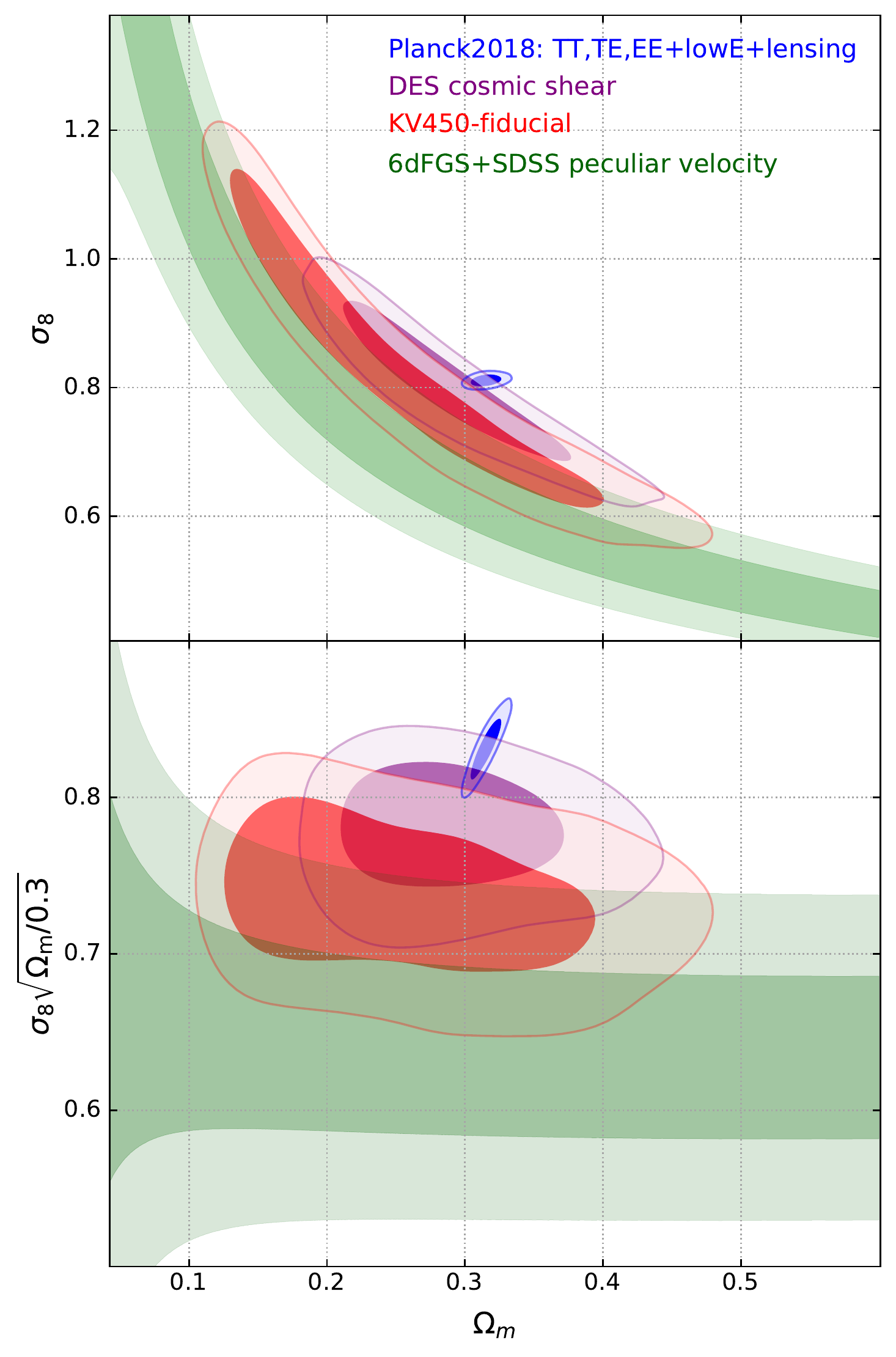}
\caption{Comparison between our $\Lambda$CDM model-dependent joint constraint on $\sigma_8$ and $\Omega_m$ (top) and on $S_8$ and $\Omega_m$ (bottom) using results from 6dFGS+SDSS peculiar velocities (this work; green), TT+TE+EE+lowE+lensing from \protect\citet[blue]{2018arXiv180706209P}, KiDS+VIKING-450 from  \protect\citet[red]{2018arXiv181206076H}, and DES cosmic shear from \protect\citet[purple]{2018PhRvD..98d3528T}. The contours show 68\% and 95\% confidence ranges.}
\label{fig_kids_450_Planck2018_sigma8}
\end{center}
\end{figure}

Figure \ref{fig_kids_450_Planck2018_sigma8} shows a comparison between our $\Lambda$CDM ($\gamma=6/11$) model dependent constraints of $\sigma_8$ shown as green confidence bands, Planck baseline results which are based on Planck TT,TE,EE+lowE+lensing \citep{2018arXiv180706209P} shown as a blue contours and the Kilo-Degree Survey (KiDS) combined with the VISTA Kilo-Degree Infrared Galaxy Survey \citep{2018arXiv181206076H}, presented by the red contours, and the Dark Energy Survey Year 1 Results based on Cosmic Shear \citep{2018PhRvD..98d3528T} plotted as a purple contours.

Our baseline $\Lambda$CDM value of $\sigma_8$ is in good agreement with the value derived by \cite{2011MNRAS.413.2906D} at comparable scale. Moreover, it agrees within $1\sigma$ with other low-redshift probes such as the Kilo Degree Survey (KiDS; \citealt{2017MNRAS.465.1454H}), and the Kilo-Degree Survey combined with the VISTA Kilo-Degree Infrared Galaxy Survey (KiDS+VIKING-450; \citealt{2018arXiv181206076H}) and within $1.5\sigma$ with galaxy clustering and weak lensing from the Dark Energy Survey (DES; \citealt{2018PhRvD..98d3526A,2018PhRvD..98d3528T}). It is also within $2\sigma$ of the Planck measurement from Sunyaev-Zeldovich cluster counts by \cite{2014A&A...571A..20P}. However, it is in $3\sigma$ tension with the latest results from \cite{2018arXiv180706209P}. 

Forthcoming surveys such as DESI \citep{2016arXiv161100036D}, Taipan \citep{2017PASA...34...47D}, WALLABY \citep{2020arXiv200207311K}, and SKA1-MID \citep{2018arXiv181102743S} will provide redshifts for millions of galaxies and true distances for several hundred thousand. They will provide the best measurements of the growth of structure by combining redshift-space distortions and direct peculiar velocity measurements (see, e.g., Figure~7 of \citealt{2017PASA...34...47D}).

\section{Summary}
In this paper, we selected a sample of 24848 elliptical galaxies from SDSS Data Release 14 ideally suited for Fundamental Plane work. We re-measured the velocity dispersion for each of those galaxies using a modified version of pPXF to implement the MC error estimation. 

We used the 3D Gaussian model to fit the Fundamental Plane. We compared our results with the previously fitted 6dFGSv Fundamental Plane. The intrinsic scatter for both relation are almost the same. However, the scatter in the $r$-direction has been improved from $29\%$ for 6dFGSv sample to $25\%$ in the SDSS sample. This improvement is mainly because of the high resolution velocity dispersion from SDSS.       

We presented a new method to simultaneously fit for the Fundamental Plane parameters as well as the velocity field parameters. We took advantage of the new imaging surveys (e.g. PS1, DES, DECaLS, VHS, and SkyMapper) to clean-up the 6dFGSv sample. we rejected $20\%$ of the
6dFGSv galaxies (i.e. 1773 out of the 8803). In addition, we applied a redshift cut of $z<0.055$ to the SDSS sample. The reason for this cut is two fold: first, it is the limit of the 2M$++$ reconstructed density field; second, it is the same redshift cut applied to the 6dFGSv sample.

Using the new method and the new samples, we fit the direct Fundamental Pane $a$, $b$, $c$, and $\sigma$ simultaneously with the velocity field parameters $\beta$ and the external bulk flow $V_{ext}$. 

We used the fitted parameters to constrain the growth rate of cosmic structure as well as deriving a model dependent $\sigma_8$. We compare our finding with other low and high redshift probes. Our measurements agree more with low redshift methods than high redshift probes.

\section*{Acknowledgments}
In this work we used the following packages: GetDist package as part of CosmoMC \citep{2002PhRvD..66j3511L,2013PhRvD..87j3529L,2019arXiv191013970L}, emcee \citep{2013PASP..125..306F}, Astropy \citep{2013A&A...558A..33A}, and Matplotlib \citep{2007CSE.....9...90H}. 

KS acknowledges funding from the Gruber Foundation as the 2017 IAU Fellow. MMC acknowledges support from Australian Research Council Discovery Projects grant DP160102075. JRL was supported by the Science and Technology Facilities Council through the Durham Astronomy Consolidated Grants ST/P000541/1 and ST/T000244/1

KS thanks M. Hashim, S. Oh, and F. D'Eugenio for many useful discussions. 

We acknowledge the effort of the staff at the Australian Astronomical Observatory who have undertaken the survey observations on the UK Schmidt telescope. Funding for the Sloan Digital Sky Survey (SDSS) has been provided by the Alfred P. Sloan Foundation, the Participating Institutions, the National Aeronautics and Space Administration, the National Science Foundation, the U.S. Department of Energy, the Japanese Monbukagakusho, and the Max Planck Society. The SDSS Web site is http://www.sdss.org/.

The SDSS is managed by the Astrophysical Research Consortium (ARC) for the Participating Institutions. The Participating Institutions are The University of Chicago, Fermilab, the Institute for Advanced Study, the Japan Participation Group, The Johns Hopkins University, Los Alamos National Laboratory, the Max-Planck-Institute for Astronomy (MPIA), the Max-Planck-Institute for Astrophysics (MPA), New Mexico State University, University of Pittsburgh, Princeton University, the United States Naval Observatory, and the University of Washington. This publication makes use of data products from the Two Micron All Sky Survey, which is a joint project of the University of Massachusetts and the Infrared Processing and Analysis Center, funded by the National Aeronautics and Space Administration and the National Science Foundation.

This research has made use of the NASA/IPAC Extragalactic Database (NED),
which is operated by the Jet Propulsion Laboratory, California Institute of Technology,
under contract with the National Aeronautics and Space Administration

\section*{Data availability}
The two main catalogues used in this article are available in the online supplementary material. All other data will be shared on reasonable request to the corresponding author.

\bibliographystyle{mn2e.bst}
\bibliography{references_n}

\appendix

\label{lastpage}

\end{document}